\documentclass[11pt]{article}

\usepackage{epsfig,amsmath,latexsym,amssymb}
\usepackage{graphicx}
\usepackage{lscape}
\usepackage{picture, eso-pic, tikz} 
\usepackage{dsfont}
\usepackage{listings}
\usepackage{changes}
\usepackage{bbding}
\usepackage{url}
\usepackage{hyperref}
\usepackage{tablefootnote}
\usepackage{booktabs}
\usepackage{algorithm}
\usepackage{tcolorbox}

\usepackage{tikz}
\usetikzlibrary{arrows,shapes,er,positioning,arrows.meta,calc}
\tikzset{
     arrow/.style = { thick,  ->, >=Triangle},
}

\renewcommand{\baselinestretch}{1.1}
\def\R{{\mathbb R}}  
\def\E{{\mathbb E}}  %

\newcommand{\Remm}[1]{}
\newtheorem{theo}{Theorem}[section]

\newtheorem{model ass}[theo]{Model Assumptions}

\newtheorem{rems}[theo]{Remarks}

\numberwithin{equation}{section}

\definecolor{MyGray}{rgb}{0.92,0.92,0.92}


\newcommand{\bl}[1]{\textcolor{blue}{{#1}}}

\newcommand{\mg}[1]{\textcolor{magenta}{{#1}}}

\definecolor{British racing}{rgb}{0.0, 0.5, 0.0}

\def\bX{\boldsymbol{X}}

\def\bvartheta{\boldsymbol{\vartheta}}

\begin{document}
\author{Ronald Richman\footnote{insureAI, ronaldrichman@gmail.com}
  \and
  Mario V.~W\"uthrich\footnote{Department of Mathematics, ETH Zurich,
mario.wuethrich@math.ethz.ch}}

\date{Version of \today}
\title{One-Shot Individual Claims Reserving}
\maketitle

\begin{abstract}
  \noindent
Individual claims reserving has not yet become established in actuarial practice. We attribute this to the absence of a satisfactory methodology: existing approaches tend to be either overly complex or insufficiently flexible and robust for practical use.
Building on the classical chain-ladder (CL) method, we introduced a new perspective on individual claims reserving in Richman and W\"uthrich [{\tt arXiv:2602.15385}]. This manuscript has sparked considerable discussion within the actuarial community. The aim of the present paper is to continue and deepen that discussion, with the ultimate goal of advancing toward a new standard for micro-level reserving.
\medskip


\end{abstract}

\section{Introduction}
We recently uploaded an individual claims reserving proposal to {\tt arXiv:2602.15385} that addressed many of the shortcomings present in the published micro-reserving literature; see Richman–Wüthrich \cite{PtU}. Starting from the classical chain-ladder (CL) method of Mack \cite{Mack}, we derived an alternative representation of the ultimate claim predictor. This alternative representation motivates a direct estimation of projection-to-ultimate (PtU) factors which allow for a {\it one-shot forecast} of the ultimate claims. The approach extends naturally to individual claims reserving for reported but not settled (RBNS) claims, allowing one to incorporate arbitrary input information (including dynamic stochastic covariates) into the estimation procedure of individual RBNS claims reserves.

Our {\tt arXiv} paper has stimulated substantial discussion of the proposed approach, indicating that this may be a promising way forward for individual claims reserving. It has also been noted that our main CL result is not new, it had previously appeared in the literature by Lorenz--Schmidt \cite{LorenzSchmidt0}. The purpose of the present manuscript is to address the many points raised in these discussions and to examine the open issues identified, we also refer to Richman–Wüthrich \cite[Section 5]{PtU} for a list of potential next steps.

\medskip

We begin by summarizing some of the feedback received; an acknowledgement is found at the end of this section.

\begin{itemize}
\item ``Maybe I am confusing the methods here but for triangles your PtU factors are called grossing-up factors, could that be? (because I think this recursive technique was already used for triangles in the grossing-up method of  (Handbook of Loss Reserving, Radtke, p 127))?''; comment by Florian Gerhardt.
\end{itemize}

This is indeed correct -- thank you! We were not aware of the corresponding result of Lorenz--Schmidt \cite[page 130]{LorenzSchmidt}; in fact, this CL result was first published by the same authors \cite{LorenzSchmidt0} in 1999. Following their theorem in Lorenz--Schmidt \cite[page 130]{LorenzSchmidt}, the authors state “..., the grossing up method is irrelevant in practice.” We believe that this assessment is too pessimistic in the era of machine learning. In our view, precisely this structural perspective provides the key to bringing individual claims reserving into practical applications. This is also supported by the following comments that we received:

\begin{itemize}
\item ``Your comment 'An alternative of building full simulation models for complex claims processes with multiple stochastic covariates and nested projections is not very practical' is interesting -- this is basically what I was working on at the start of my career! Admittedly without a lot of the complexity you could build into it, because it was too computationally intensive back then. But I’m not at all convinced it’s impractical nowadays.''; comment by Chris Dolman.
\item ``I experimented with a similar idea several years ago and a couple of clients reported back that it worked well. They kept their existing triangle methods to keep the regulator happy but used the individual claim ultimate cost machine learning approach to inform the triangle parameters.''; comment by Colin Priest.
\item ``I've worked on something very similar two years ago. I called it 'Similarity score weighted micro chain ladder'. It also had the same recursive pattern, but predictions of the point to ultimate estimates per claim were made with simple classification techniques to estimate distance measures between the claim to predict with older claims to take a weighted average point to ultimate (as opposed to using a NN for this as in your case). It becomes computationally complex really quickly. One thing I found quite useful for computational efficiency here was to group claims in operational time bands instead of development periods. This works especially well if you have transaction dates in your data.
In this work we were specifically looking to close the gap between micro reserving and triangle reserving since adoption of micro reserving has been so poor.  Great to see you guys also moving into this space.''; comment by Stephan Marais.
\item ``Wondering if it’s the same idea as in the addendum [to Semenovich \cite{Semenovich}]?
This is something I looked into originally 10+ years ago. But never got to a satisfactory formulation around doing joint IBNR and IBNER estimation in a single regression model until the method in the addendum.''; comment by Dimitri Semenovich.
\item ``Did something similar to estimate prob default after 12 months in IFRS9 ...''; comment by Willem Ras.
  \item ``This is really interesting work. I think micro reserving is the next step in reserving, not just for improved accuracy, but also for greater flexibility. As soon as data is compressed into a triangle, much of the claim-level information is lost. Parodi’s triangle-free paper \cite{Parodi} makes this point nicely with the analogy of moving from a high-resolution to a low-resolution picture.
Working at claim level makes it easier to reflect environmental shifts, such as inflation shocks, detect changes in business mix earlier or generate different views since IBNR is produced from bottom-up.
But triangles have survived for nearly a century for a reason: they are simple, robust, and hard to fool. So I expect they will stick around for the foreseeable future.''; comment by Claudio Rebelo.
\end{itemize}

The above feedback is very exciting, and it seems that many colleagues have been considering such or a similar approach. We see our first main contribution in documenting all these similar thoughts and in making the link to the PtU factors in a CL context, which turned out to be the grossing-up method of Lorenz--Schmidt \cite{LorenzSchmidt}. Concerning the last remark, (individual) claims reserving data has a natural triangular structure through censoring at the present evaluation date. The present paper illustrates how triangular methods on aggregated data can be refined to be able to operate on individual claim-level information.

There were two main critical points raised above: the computational side (which indeed may be demanding) and the recursive structure (which may be prone to biases). We will come back to these issues in the next bullet points and in our numerical studies below. The proposal in the addendum to Semenovich \cite{Semenovich} starts from a cross-classified Poisson model, which provides a different way of computing the CL reserves; see Hachemeister--Stanard \cite{HachemeisterStanard}, Kremer \cite{Kremer} and Mack \cite{Mack1991}. This cross-classified structure presents a more restrictive model from a mathematical perspective, but from a computational viewpoint it circumvents the recursive estimation and forecast structure. This indeed may provide another very promising alternative, i.e., by solving the model estimation by a single maximum likelihood estimation (MLE) procedure.

\begin{itemize}
\item ``The approach is very interesting and promising, especially, because it does not propose another ML approach, but it is rather thinking about restructuring the data to preform individual claims reserving.''; comment by Christian Lorentzen.
\item  ``Why don't you start with a GLM instead of the neural network?''; comment by Christian Lorentzen.
  \item ``Sounds interesting. Certainly jumping to ultimate is more interesting and aligned with UW views on variability. Does the model allow one / $n$-time step forward projections? Helps S2 reporters with MVM calcs.''; comment by David Menezes.
\item ``Curious what the advantage is over just building a GBM that samples over the future space for the underlying data and uses the length of the forecast horizon as one of the inputs?''; comment by Alex Rowley.
\item ``Nice approach for RBNS under CL! Being a fan of generic lightweight neural models, I recommend exploring an additional pathway, further relaxing model assumptions: the use of a (neural) continuous time-to-event framework including interval censoring. In this context IBNR and RBNS can be interpreted as intermediate states between claim occurrence and final settlement. The entire stochastic claim process then becomes fully explicit, with simulation as the tool for deriving all estimates.''; comment by Anne van der Scheer.
\end{itemize}

The first item of the above list is a perfect summary of our intention, i.e., it is not about a specific model architecture, but rather about how to organize the data. The second item, starting with a generalized linear model (GLM), is an excellent proposal that we should already have considered in our first paper \cite{PtU}. Ultimately, any reasonable regression model may work, the specific choice will depend on its purpose, see Shmueli \cite{Shmueli} on `To explain or to predict?', e.g., for cash flow forecasting or mid-year reserving transformer decoders could be useful tools. However, the proposal of starting with a GLM is a very valid one, and one of the exciting findings of the present case study is that even a linear regression model does an excellent job! The linear regression can be computed very efficiently, and therefore, we can even build on an individual claims bootstrap algorithm here to assess model uncertainty.

If we understand correctly, the last two items of the above list are related to the addendum of Semenovich \cite{Semenovich} who proposed a cross-classified Poisson model that can simultaneously deal with incurred but not reported (IBNR) and RBNS claims. Using the cross-classified Poisson structure, the problem can be solved in closed-form using MLE. For more complex architectures, this seems less clear. As explained in our previous paper \cite{PtU}, we rather prefer to circumvent a simulation extrapolation as this is a topic with its own difficulties. Adding the length of the forecast horizon may be an interesting proposal to shrink the number of necessary regression models. However, at the current stage, it seems not fully aligned with our recursive structure of estimating the PtU factors.

\begin{itemize}
\item ``One model per accident period seems a lot.'' 
\item ``Can this also be used for quarterly (mid-year) reserving?''
\item ``It would be nice to have a simple IBNR model to be able to compare the results to classical CL.''
\end{itemize}

We agree that computationally one model per accident/development period can be demanding. However, this is not any different from the CL method because each CL factor needs to be interpreted as 'one model' in this set-up: note that each CL factor solves a regression problem (without an intercept). In the present paper, we solve everything with linear regression models which can be computed very efficiently.

It allows for quarterly reserving. In fact, the input can be in continuous time, even if the prediction is only on an annual grid. After year 2000, when many insurance companies transitioned from annual to quarterly reporting, they initially used a grossing-up method to complete a partially observed calendar year to receive an end-of-year forecast. Based on this end-of-year forecast they performed a CL or Bornhuetter--Ferguson \cite{BF} method on an annual grid.

Finally, incurred but not reported (IBNR) claim forecasting is a crucial missing piece in our previous work, Richman--W\"uthrich \cite{PtU}, which we are going to tackle in Section \ref{sec: IBNR reserving}, below.

\bigskip

{\bf Organization of this manuscript.}
\begin{itemize}
  \item
Section \ref{sec: Chain-ladder method} revisits the classic CL method. We discuss the transition from the iterative one-period ahead roll-forward extrapolation method to recursive one-shot ultimate claim prediction using the PtU factors. Typically, this is done on aggregated cumulative payments, and we explain its decomposition to individual claims observations. This paves the path to bootstrapping individual claims histories, and we challenge Mack's \cite{Mack} model error estimate by a corresponding individual claims history bootstrap analysis.
\item In Section \ref{Chain-ladder RBNS reserving}, we distinguish claims according to their reporting status  -- resulting in RBNS and IBNR claims. This is a crucial step in individual claims reserving to ensure that PtU factors are estimated on consistent claims cohorts -- this is the first step that significantly differs from CL reserving on aggregated claims, and it is the crucial step to prepare for individual claims reserving. This step also provides a novel decomposition of the classical CL reserves into RBNS reserves and IBNR reserves.
\item Section \ref{Individual ultimate prediction using machine learning} is our core section. We dive into individual claims reserving for RBNS claims, and interestingly, we see that a linear regression model on the individual claim features can attain an excellent predictive performance.
  \begin{itemize}
  \item Section \ref{Recursive individual RBNS claims reserving} presents the generic recursive one-shot PtU forecast algorithm for RBNS claims. This is our core tool for individual claims reserving; see Algorithm \ref{algorithm: PtU one-shot}.
  \item Section \ref{Lab: Accident insurance example -- linear regression} gives the first real data application of Algorithm \ref{algorithm: PtU one-shot}. This application is fully based on linear regression models (and a Markov assumption).
  \item Section \ref{Lab: Linear regression bootstrap results} applies an individual claims history bootstrap to the previous individual claims reserving method. This can be done efficiently because all predictive models are based on linear regressions.
    \item Section \ref{Lab: Accident insurance example -- feed-forward neural network} challenges the linear regression models with neural networks, with the result that the networks do not provide a significantly better predictive result.
  \item Section \ref{Transformer architecture} analyzes transformer architectures to see whether we can gain predictive power by inputting the entire past claims history (by dropping the Markov assumption). In our small-scale example, the answer is negative, but this should be reconsidered on bigger datasets to receive better answers, i.e., this section rather provides a proof of concept in the sense that transformers can be integrated into the forecast procedure, and they provide stable results.
  \end{itemize}
  
\item In Section \ref{The role of claims incurred}, we analyze the predictive power of claims incurred information. Our finding is that on a (small) liability insurance dataset, the claims incurred information gives more accurate forecasts than the individual cumulative payment information, in particular, in combination with the claims status information.
\item In Section \ref{sec: IBNR reserving}, we discuss setting the IBNR reserves for late reported claims. This is performed by a simple CL application on the predicted ultimates of RBNS claims.
  \item Section \ref{Conclusions and Outlook} concludes and gives an outlook.
  \end{itemize}

\medskip

{\bf Acknowledgement.} Thank you very much for the numerous and very useful feedback (in alphabetical order):
Chris Dolman,
Florian Gerhardt,
Syed Kirmani,
Christian Lorentzen,
Stephan Marais,
David Menezes,
Colin Priest,
Willem Ras,
Claudio Rebelo,
Alex Rowley,
Dimitri Semenovich, 
Anne van der Scheer.

\section{Chain-ladder method - revisited}
\label{sec: Chain-ladder method}
We begin by revisiting Mack's \cite{Mack} CL algorithm and its reformulation that leads to the appealing structure for individual claims reserving using machine learning (ML) methods. This gives us the motivation and the basis for all subsequent derivations; for full technical details we refer to Richman–Wüthrich \cite{PtU}.

\subsection{Chain-ladder algorithm - recursive one-shot forecast}
\label{Chain-ladder algorithm - recursive one-shot forecast}
\begin{tcolorbox}[title=] 
This section presents the step going from the one-period ahead roll-forward CL extrapolation to the recursive one-shot ultimate claim forecast. For this we define the PtU factor that allows one to gross-up the last observed cumulative payments.
\end{tcolorbox}

We consider $I$ accident periods and a maximal development delay $J$, throughout $J<I$.
Cumulative payments for the claims in accident period $i \in \{1,\ldots, I\}$ at development delay $j \in \{0,\ldots, J\}$ are denoted by $C_{i,j}$, and we assume that these cumulative payments are strictly positive for all indexes $(i,j)$; cumulative payments $C_{i,j}$ means that these variables collect all the payments made for accident year  $i$ within the development periods up to period $j$.

\begin{figure}[htb!]
\begin{center}
\includegraphics[width=\textwidth]{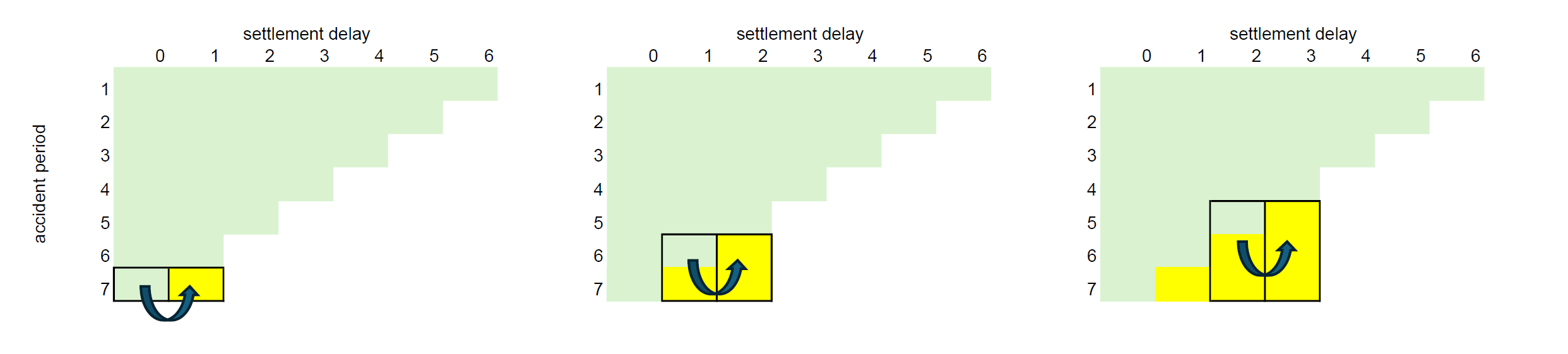}
\end{center}
\caption{One-period ahead roll-forward extrapolation to predict the ultimate claims $C_{i,J}$ using the observations $C_{i,I-i}$, $i>I-J$, at time $I$ (for $I=7$ and $J=6$); this figure is taken from \cite{PtU}.}
\label{fig:CL1}
\end{figure}

\begin{figure}[htb!]
\begin{center}
\includegraphics[width=\textwidth]{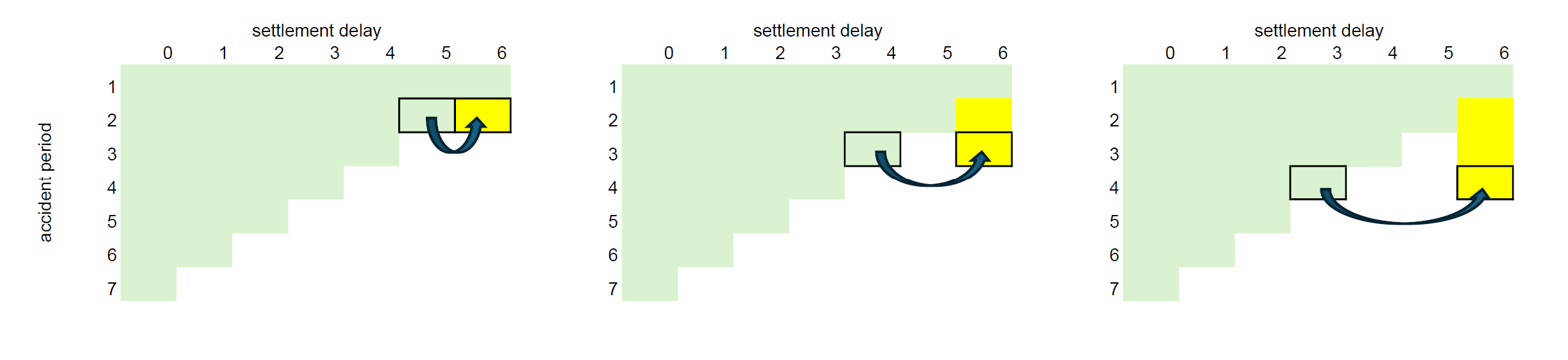}
\end{center}
\caption{Backward (in time) one-shot predictions of the ultimate claims $C_{i,J}$, $i>I-J$, using the `directly estimated' PtU factors $(\widehat{F}^{\rm CL}_j)_{j=0}^{J-1}$ given in  \eqref{PtU CL factor}: (left-middle-right) correspond to $j-1=J-1=5$, $j-1=4$ and $j-1=3$; this figure is taken from \cite{PtU}.}
\label{fig:CL2}
\end{figure}

At calendar time $I$, we have observed the upper triangle/trapezoid
\begin{equation}\label{available data at time I}
{\cal D}_I=
\left\{ C_{i,j};~   i+j \le I,\, 1\le i \le I,\, 0\le j \le J \right\},
\end{equation}
this corresponds to the green triangles in Figures \ref{fig:CL1} and \ref{fig:CL2}.
The general goal is to predict the ultimate claims $C_{i,J}$ for all accident periods $i$ with $i+J>I$, i.e., the accident periods that are not fully developed at time $I$.

For the CL reserving method, we estimate the so-called {\it CL factors} $(f_j)_{j=0}^{J-1}$ at time $I$ by
\begin{equation}\label{CL factor estimates}
\widehat{f}^{\rm CL}_j = \frac{\sum_{i=1}^{I-(j+1)}C_{i,j+1}}
{\sum_{i=1}^{I-(j+1)}C_{i,j}}.
\end{equation}
The {\it CL predictors} at time $I$ of the ultimate claims for accident periods $i>I-J$ are defined by
\begin{equation}\label{forward path}
\widehat{C}^{\rm CL}_{i,J}=
C_{i,I-i}\prod_{j=I-i}^{J-1} \widehat{f}^{\rm CL}_j;
\end{equation}
these are the classic CL predictors; see Mack \cite{Mack}. Define the {\it projection-to-ultimate} (PtU) {\it factors}
\begin{equation}\label{CL PtU}
  \widehat{F}^{\rm CL}_j = \prod_{l=j}^{J-1} \widehat{f}^{\rm CL}_l \qquad \text{ for $j\in \{0,\ldots, J-1\}$.}
\end{equation}
These give the identical CL predictors for $i>I-J$
\begin{equation}\label{grossing-up method}
\widehat{C}^{\rm CL}_{i,J}=
C_{i,I-i}\prod_{j=I-i}^{J-1} \widehat{f}^{\rm CL}_j = C_{i,I-i}\,\widehat{F}^{\rm CL}_{I-i}.
\end{equation}
In the actuarial literature, the PtU factors \eqref{CL PtU} are also called {\it grossing-up factors}, making the reserving method in \eqref{grossing-up method} a grossing-up reserving method; see Lorenz--Schmidt \cite{LorenzSchmidt}.

The mechanics of the CL estimation and prediction procedure \eqref{forward path} is illustrated in
Figure \ref{fig:CL1}. It has the following iterative one-period ahead roll-forward structure
\begin{equation*}
\widehat{C}^{\rm CL}_{i,J}~=~C_{i,I-i}
\prod_{j=I-i}^{J-1} \widehat{f}^{\rm CL}_j~=~
\underbrace{\underbrace{C_{i,I-i}  \cdot \widehat{f}^{\rm CL}_{I-i}}_{I-i ~\to~ I-i+1}\,\cdot \,\widehat{f}^{\rm CL}_{I-i+1}}_{I-i ~\to~ I-i+2} \,\cdot\, \ldots \,\cdot \,\widehat{f}^{\rm CL}_{J-1}.
\end{equation*}

It is precisely this iterative one-period ahead roll-forward extrapolation structure that poses significant difficulties in individual claims reserving using ML methods because dealing with such extrapolations of stochastic processes is generally a difficult problem.
This has led to the idea of trying to perform a direct {\it one-shot forecast} of the ultimate claim by {\it directly estimating} the PtU factor, where 'directly estimating' means that we do {\it not} go through the iterative one-period ahead construction \eqref{CL PtU}, but we directly estimate the PtU factor in a single computation, see \eqref{PtU CL factor}.
As proved in  
Lorenz--Schmidt \cite{LorenzSchmidt} and verified in 
Richman–Wüthrich \cite[Proposition 2.2]{PtU}, this is possible. 
Algorithm \ref{algorithm: recursive CL one-shot} gives this one-shot prediction variant
of the CL predictors \eqref{forward path};
for mathematical details see Richman–Wüthrich \cite[Proposition 2.2]{PtU}, and it is illustrated in Figure \ref{fig:CL2}.

\begin{algorithm}
\caption{Recursive one-shot CL prediction algorithm.}\label{algorithm: recursive CL one-shot}
\begin{itemize}
\item[(a)]{\it Initialization for $j=J$.}
For the fully settled accident periods $i \in \{1,\ldots, I-J\}$, initialize the algorithm by $\widehat{C}^{\rm CL}_{i,J}=C_{i,J}$.
\item[(b)]{\it Iteration $j \to j-1 \ge 0$.} Compute recursively
\begin{equation}\label{PtU CL factor}
\widehat{F}^{\rm CL}_{j-1}= \frac{\sum_{i=1}^{I-j}\widehat{C}^{\rm CL}_{i,J}}
{\sum_{i=1}^{I-j}C_{i,j-1}}
\qquad \text{ and } \qquad \widehat{C}^{\rm CL}_{I-(j-1),J}=C_{I-(j-1),j-1}\,\widehat{F}^{\rm CL}_{j-1}.
\end{equation}
\end{itemize}
\end{algorithm}

Remark, the predictors \eqref{PtU CL factor} and  \eqref{forward path} are identical; see 
Richman–Wüthrich \cite[Proposition 2.2]{PtU}. That is, 
\eqref{PtU CL factor} gives a different representation of 
\eqref{forward path} which is more appealing in individual claims reserving.

\subsection{Individual claims - available data}
\begin{tcolorbox}[title=]
This section introduces the individual claims and their individual claim histories. We distinguish between cumulative claims and aggregated claims, and we explain the difference between RBNS and IBNR claims.
\end{tcolorbox}

The cumulative payments $C_{i,j}$ consider aggregated payments over {\it all} claims having occurred in accident period $i$ up to development period $j$. We emphasize that we distinguish the meanings 
\begin{itemize}
\item  of {\it cumulative} referring to summing payments over development periods $j$, and 
\item of {\it aggregated} referring to summing over different individual claims.
\end{itemize}

We now shift our focus to individual claims modeling.
Assume there are $N_i$ claims that occurred in accident period $i$. We label these claims by $\nu =1, \ldots, N_i$, and we study each of these claims individually.
Denote the {\it reporting delay} of the $\nu$-th claim of accident period $i$ by $T_{i|\nu} \ge 0$; the reporting delay is the time difference between the occurrence period $i$ of the claim and its reporting (notification) period $i+T_{i|\nu}$ at the insurance company. Thus, after reporting delay $j$, all claims $\nu$ with reporting delay $T_{i|\nu}\le j$ are reported at the insurance company, and the claims $\nu$ with $T_{i|\nu}> j$ are not reported at time $i+j$.

\medskip

For a fixed time point $I$, called {\it evaluation date}, we have the following two classes of claims:
\begin{itemize}
\item we call the claims that are not reported yet, $i+T_{i|\nu}> I$, {\it incurred but not reported} (IBNR) claims, and 
\item all other claims, $i+T_{i|\nu}\le I$, are called {\it reported but not settled} (RBNS) claims. By convention, RBNS claims include {\it all} reported claims, these can be open or closed (settled), as some closed claims may require a re-opening due to late unexpected further claim developments.
\end{itemize}
As soon as a claim $\nu$ is reported (RBNS), the insurance company starts to collect information about this specific claim. E.g., the insurance company can study its {\it individual cumulative payment process} given by
\begin{equation}\label{individual claims payment process}
C_{i, 0:J| \nu} = \left[C_{i,0|\nu}\, \mathds{1}_{\{T_{i|\nu} \le 0\}}, \,C_{i,1|\nu} \,\mathds{1}_{\{T_{i|\nu} \le 1\}}, \ldots, \,C_{i,J|\nu}\, \mathds{1}_{\{T_{i|\nu} \le J\}}\right].
\end{equation}
We {\it mask} $C_{i,j|\nu}=0$ all IBNR periods $j<T_{i|\nu}$, i.e., before the claim has been reported to the insurance company; one could also use any other mask value. A lower index $_{0:J}$ generically denotes a sequence that considers the time indexes $j=0,\ldots, J$.

\begin{figure}[htb!]
\begin{center}
\begin{minipage}[t]{0.45\textwidth}
\begin{center}
\includegraphics[width=\textwidth]{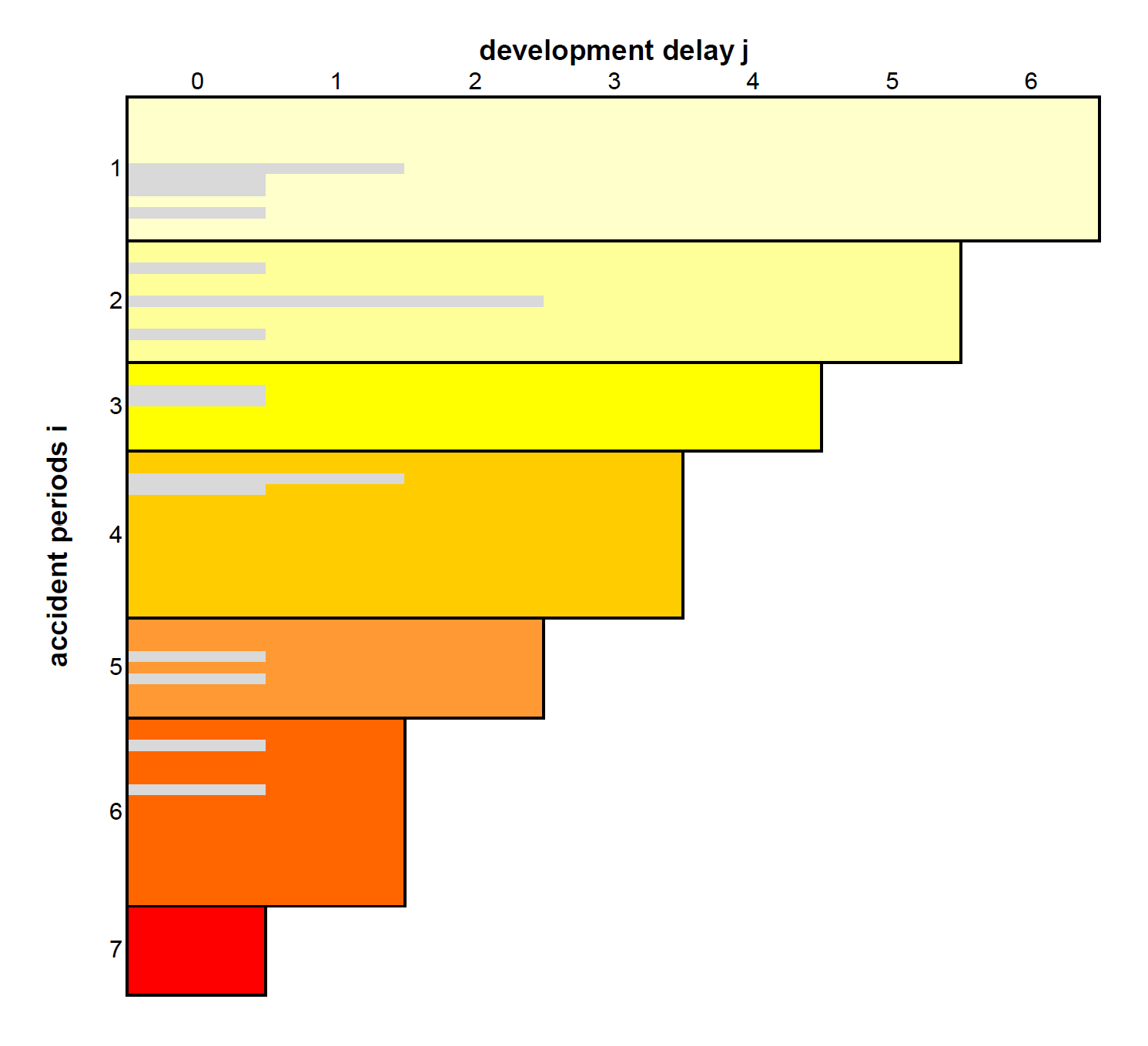}
\end{center}
\end{minipage}
\hspace{1cm}
\begin{minipage}[t]{0.45\textwidth}
\vspace{-5cm}
\begin{center}
\includegraphics[width=\textwidth]{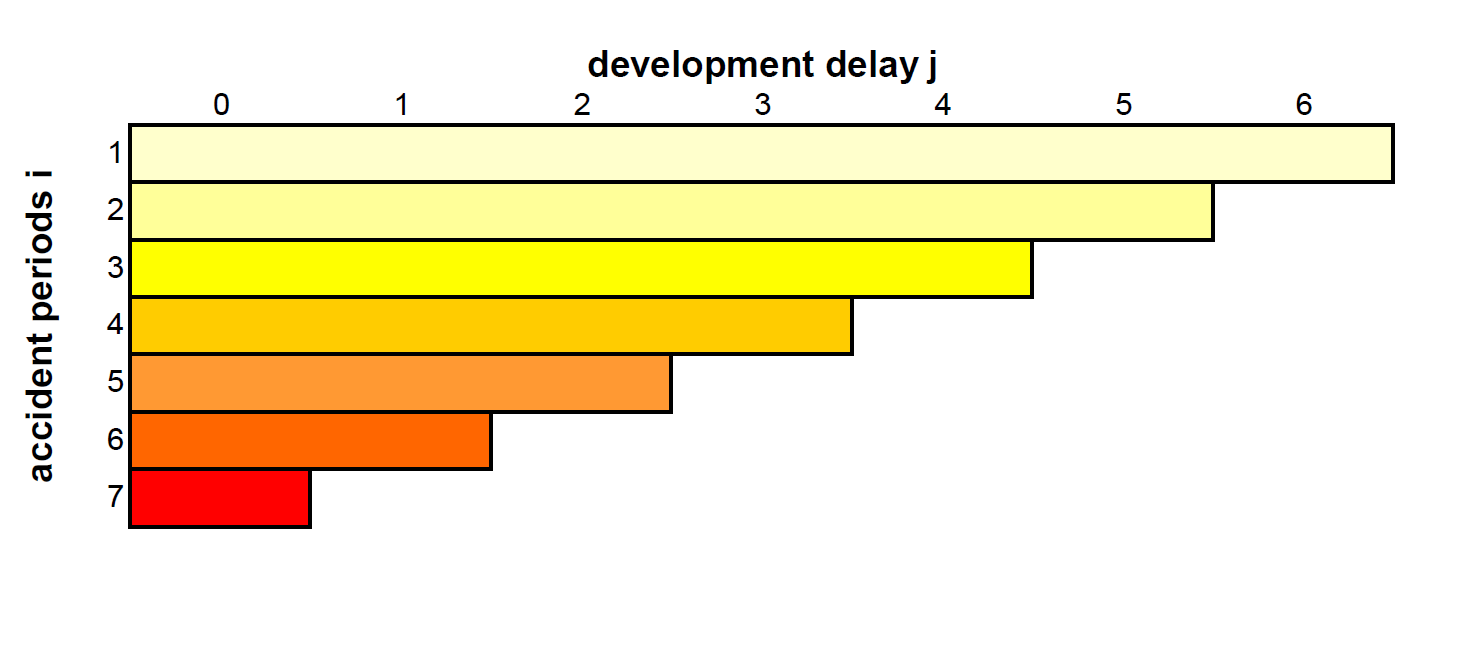}
\end{center}
\end{minipage}
\end{center}
\caption{(lhs) Individual cumulative payments $C_{i,j|\nu}$ in the upper triangle $i+j \le I$ (each row is one claim, period $i=4$ has twice as many claims as $i=3$), and (rhs) aggregated cumulative claims $C_{i,j}$ in the upper triangle. Late reportings are illustrated by gray bars in the left-hand side figure.}
\label{Fig: individual claims}
\end{figure}

The {\it aggregated} cumulative payments $C_{i,j}$ over all claims that have occurred in accident period $i$ up to development period $j$ are then computed by
\begin{equation}\label{individual to aggregate}
C_{i,j} = \sum_{\nu = 1}^{N_i} 
C_{i,j|\nu}\, \mathds{1}_{\{T_{i|\nu} \le j\}}
= \sum_{\nu:\, T_{i|\nu} \le j} C_{i,j|\nu},
\end{equation}
we are going to use the latter notation as it is more convenient. 
Naturally, but importantly for the further understanding, only RBNS claims can have payments, this motivates the expression $C_{i,j|\nu}\, \mathds{1}_{\{T_{i|\nu} \le j\}}$ in \eqref{individual claims payment process} and \eqref{individual to aggregate}.

 Figure \ref{Fig: individual claims} (lhs) indicates individual cumulative payment histories $C_{i,0|\nu}, \ldots, C_{i,I-i|\nu}$ in the (observed) upper triangle -- each row corresponds to one claim. The gray bars show late reported claims, e.g., in the first accident period $i=1$, there is one claim with reporting delay $T_{1|\nu}=2$. Such claims with a reporting lag of 2 periods are missing for the most recent accident periods $i=6,7$, because they are not reported yet, i.e., they are IBNR claims at the evaluation date $I=7$. The right-hand side of Figure \ref{Fig: individual claims} shows its aggregated version $C_{i,0}, \ldots, C_{i,I-i}$, see \eqref{individual to aggregate}, where all the payments are aggregated within accident periods $i$ and development periods $j$.
 
\medskip

The individual payment information \eqref{individual claims payment process} is sufficient to compute the CL predictors \eqref{forward path}. However, often there is additional individual claim information available. We denote the process of the additional individual information by
\begin{equation}\label{covariates individual process}
\bX_{i, 0:J| \nu} = \left[\bX_{i,0|\nu}\, \mathds{1}_{\{T_{i|\nu} \le 0\}}, \,\bX_{i,1|\nu} \,\mathds{1}_{\{T_{i|\nu} \le 1\}}, \ldots, \,\bX_{i,J|\nu}\, \mathds{1}_{\{T_{i|\nu} \le J\}}\right],
\end{equation}
where we again use a mask for $\bX_{i,j|\nu}$ for all IBNR periods $j< T_{i|\nu}$, this corresponds to the gray bars in Figure \ref{Fig: individual claims} (lhs). Thus, each claim $\nu=1,\ldots, N_i$ of accident period $i$ is described by a {\it claim settlement process} ({\it individual claim history})
\begin{equation}\label{definition claims settlement process}
{\cal C}_{i|\nu}=(C_{i,0:J|\nu}, \bX_{i,0:J|\nu}).
\end{equation}
The claim settlement components $(C_{i,j|\nu}, \bX_{i,j|\nu})$ before reporting  $j<T_{i|\nu}$ are masked as IBNR periods, and at the evaluation date $I$ the entries with indexes $i+j>I$ have not been observed yet, because they lie in the future at time $I$ (this is the lower (white) triangle in Figure \ref{Fig: individual claims}).

The additional claim features collect any information about the individual claim, e.g.,
\begin{equation}\label{claims features}
\bX_{i,j|\nu}
= \begin{pmatrix}
\text{reporting delay $T_{i|\nu}$}\\
\text{business line}\\
\text{claims type}\\
\text{settlement delay $j$}\\
\text{claim status closed/open at delay $j$}\\
\text{claims incurred at delay $j$}\\
\text{case reserves at delay $j$}
\end{pmatrix}
.
\end{equation}
The first three entries are {\it static covariates} that become available at reporting, the fourth component is a {\it deterministic dynamic covariate} (it is dynamic but perfectly predictable), and the last three entries are  {\it stochastic dynamic covariates}. The information in \eqref{claims features} is called tabular, because it considers structured data that has a tabular form. However, the algorithms presented below can also deal with unstructured data, e.g., we could include a medical report into 
$\bX_{i,j|\nu}$ -- medical reports are also of stochastic dynamic nature.

For the CL method, we started from a fixed time grid, e.g., a monthly, quarterly or an annual grid, with accident period index $i$ and development delay index $j$ living on that grid. The algorithms presented below can also deal with continuous time inputs. In that case, we replace the discrete time version
\eqref{definition claims settlement process} by 
\begin{equation}\label{continuous time input}
{\cal C}_{i|\nu}=(C_{i,t|\nu}, \bX_{i,t|\nu})_{t \in [0, J]},
\end{equation}
that is, we keep a discrete time grid for the accident period $i$, but the claim settlement process lives in continuous time $t \in [0,J]$. We keep the discrete time in the accident period $i$ because the algorithms will be recursive in that time index.

\subsection{Chain-ladder algorithm on individual claims}
\begin{tcolorbox}[title=]
  This section discusses the computation of the CL factors being ratios of two claim cohorts that are not fully consistent. This precisely motivates the step going from total claims reserves to RBNS claims reserves.
  Moreover, we represent the CL factor computation as a minimization problem, which is the key to lift the CL factors to regression functions.
\end{tcolorbox}

The CL algorithm has been computed on aggregated cumulative payments $C_{i,j}$. Naturally, we can perform the same computations on individual claims. 
In view of \eqref{individual to aggregate}, the CL factors computed in formula \eqref{CL factor estimates} are equally obtained by
\begin{equation}\label{CL factors alternative}
\widehat{f}^{\rm CL}_j = \frac{\sum_{i=1}^{I-(j+1)}\sum_{\nu:\, \mg{T_{i|\nu} \le j+1}}C_{i,j+1|\nu}}
{\sum_{i=1}^{I-(j+1)}\sum_{\nu:\, \mg{T_{i|\nu} \le j}}C_{i,j|\nu}}.
\end{equation}
There are two points being worth to be raised in this alternative representation. These two points are going to be crucial for our further discussion and understanding.

\medskip

(1) The first point is that the nominator and the numerator of \eqref{CL factors alternative} do not consider the identical claim cohorts. The difference precisely concerns the claims with reporting delay $T_{i|\nu}= j+1$. This can be seen as follows
\begin{equation}\label{CL factors alternative 2}
  \widehat{f}^{\rm CL}_j = \frac{\sum_{i=1}^{I-(j+1)}\sum_{\nu:\, \mg{T_{i|\nu} \le j}}C_{i,j+1|\nu}
  \,+\, \sum_{i=1}^{I-(j+1)}\sum_{\nu:\, \mg{T_{i|\nu}= j+1}}C_{i,j+1|\nu}}
{\sum_{i=1}^{I-(j+1)}\sum_{\nu:\, \mg{T_{i|\nu} \le j}}C_{i,j|\nu}}.
\end{equation}
That is, the CL factors include a margin for late reported (IBNR) claims --
second term in the numerator of \eqref{CL factors alternative 2} --
and therefore these factors cannot serve as predictors on individual RBNS claims because they will lead to biased estimates on these individual RBNS claims, the total bias being of the size of the predicted IBNR claims. To properly account for this, we are going to modify the CL method in Section \ref{Chain-ladder RBNS reserving}, below.

\begin{figure}[htb!]
\begin{center}
\begin{minipage}[t]{0.45\textwidth}
\begin{center}
\includegraphics[width=\textwidth]{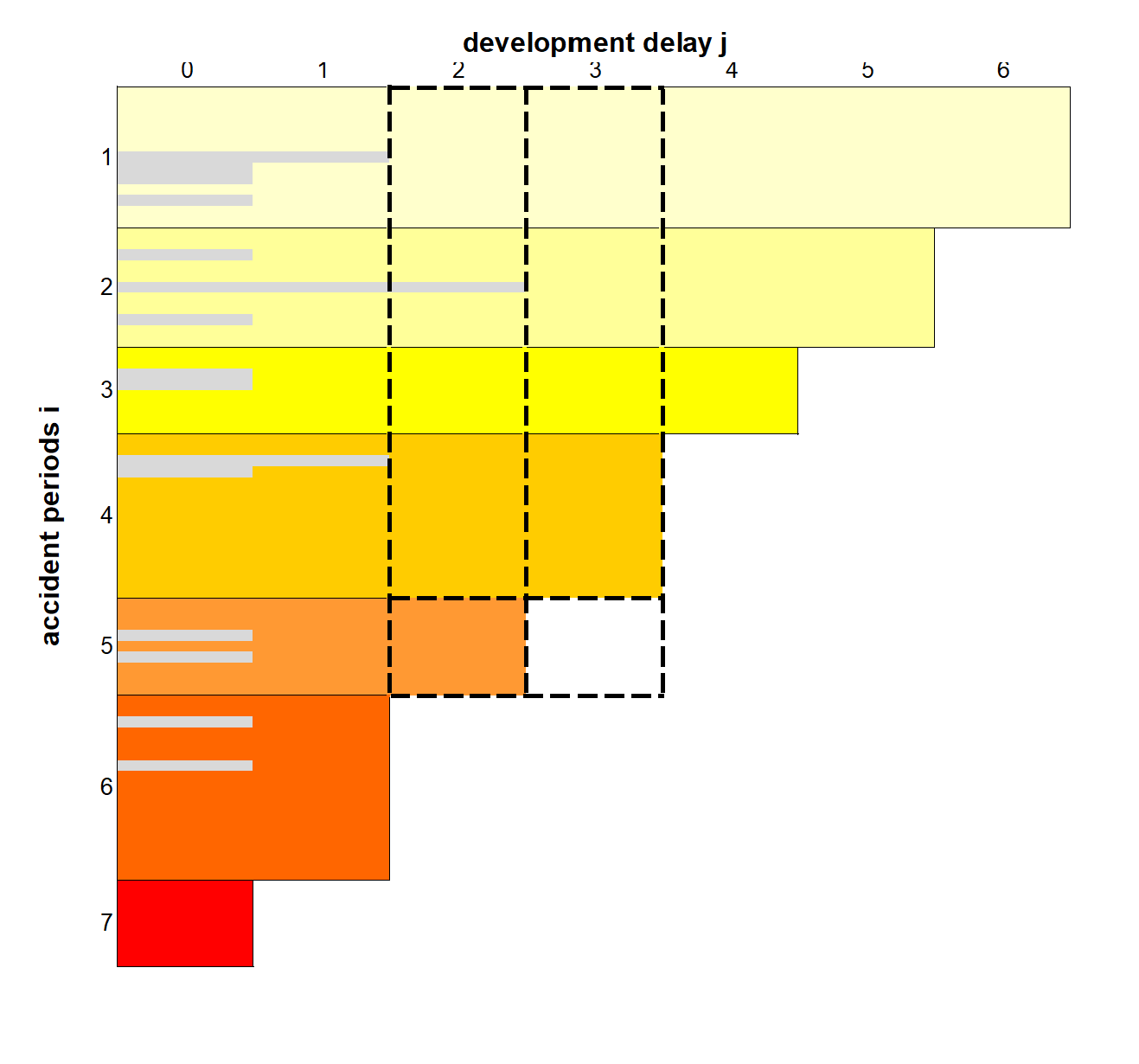}
\end{center}
\end{minipage}
\end{center}
\caption{Individual RBNS vs.~IBNR projection.}
\label{Fig: individual claims RBNS}
\end{figure}

Figure \ref{Fig: individual claims RBNS} illustrates this issue for development step $j=2 \to j+1=3$.
If we want to extrapolate the RBNS claims of accident period $i=5$, the CL factor should only contain the ratio
of claims that have been reported at settlement delay $j=2$. In Figure \ref{Fig: individual claims RBNS}, this is not the case: there is one claim $\nu$ of accident period $i=2$ with reporting delay $T_{i|\nu}=j+1=3$ (gray bar). Therefore, the columns for $j=2$ and $j+1=3$ do not contain the identical claims, and the corresponding CL ratio \eqref{CL factors alternative 2} accounts for this IBNR claim as well.

\medskip

(2) The second point we want to emphasize is that the estimator \eqref{CL factors alternative}
can be received as the solution of a weighted square minimization problem. On aggregated claims, this is related to the variance assumption in Mack's \cite{Mack} distribution-free CL model. On individual claims, this needs some care. At the moment, \eqref{CL factors alternative 2} contains IBNR claims at development delay $j$ being masked by zero, but it also contains RBNS claims that may have individual cumulative payments $C_{i,j|\nu}\ge 0$ that are equal zero. That is, on aggregated cumulative claims $C_{i,j}$ we have made the assumption of strict positivity, see Section \ref{Chain-ladder algorithm - recursive one-shot forecast}, but on individual cumulative claims $C_{i,j|\nu}$ we do not want to make this assumption as for quite some of these claims the payments may only occur later. To cope with this problem in the estimation procedure, we
select a small positive constant $\epsilon>0$, and consider the weighted square minimization problem
\begin{equation}\label{weighted square loss}
\widehat{f}^{\epsilon}_j=\underset{f_j} {\arg\min} \left\{
\sum_{i=1}^{I-(j+1)}\sum_{\nu:\, T_{i|\nu} \le j+1}
\max\{C_{i,j|\nu},\epsilon\}\left(\frac{C_{i,j+1|\nu}}{\max\{C_{i,j|\nu},\epsilon\}}-f_j \right)^2 \right\},
\end{equation}
where we impute $\epsilon$ for non-positive and IBNR claims $T_{i|\nu}=j+1$ at settlement delay $j$. 
The solution to \eqref{weighted square loss} is given by
\begin{equation}\label{epsilon limit}
\widehat{f}^{\epsilon}_j = 
\frac{\sum_{i=1}^{I-(j+1)}\sum_{\nu:\, T_{i|\nu} \le j+1}C_{i,j+1|\nu}}
{\sum_{i=1}^{I-(j+1)}\sum_{\nu:\, T_{i|\nu} \le j+1}\max\{C_{i,j|\nu},\epsilon\}}
~\le ~ \widehat{f}^{\rm CL}_j
\qquad \text{ with }\qquad 
\widehat{f}^{\epsilon}_j \uparrow \widehat{f}^{\rm CL}_j
\text{~ for  $\epsilon \downarrow 0$.}
\end{equation}
This shows that we can estimate the CL factors from minimization problems. This observation is the key to lift CL reserving to ML methods, namely, $f_j=f_j(C_{i,0:j|\nu}, \bX_{i,0:j|\nu})$ in \eqref{weighted square loss} can be made dependent on claim covariates
$(C_{i,0:j|\nu}, \bX_{i,0:j|\nu})$, which opens to door for regression modeling of the CL factors.
Below, we are going to modify this in three ways:
\begin{itemize}
\item[(1)] We will use the one-shot ultimate claim forecast variant as outlined in Algorithm \ref{algorithm: recursive CL one-shot}. This will avoid complicated iterative one-step ahead extrapolations.
\item[(2)] We will ensure that the claim cohorts considered in the nominator and numerator in \eqref{CL factors alternative} are identical, so that the method is suitable for individual RBNS claim prediction without adding a margin for IBNR claims. IBNR claims require a separate treatment.
\item[(3)] We will consider alternative objective functions because optimizing \eqref{weighted square loss} on individual claims and for flexible regression functions $f_j(C_{i,0:j|\nu},\bX_{i,0:j|\nu})$ may result in stability issues, caused by non-positive (or small) individual cumulative claims  $C_{i,j|\nu}$. 
\end{itemize}

{\bf Remarks.} Making the CL factors $f_j(C_{i,0:j|\nu},\bX_{i,0:j|\nu})$ covariate-dependent has already been considered in W\"uthrich \cite{CLnet}. This reference used a one-step ahead roll-forward extrapolation similar to \eqref{forward path}, resulting in the same difficulties as many other proposed methods in the literature. Another interesting variant is that we could replace the weights of IBNR claims in \eqref{weighted square loss} by (premium) exposures giving us a type of incremental loss ratio method for IBNR claims.

\subsection{Lab: Chain-ladder reserving and individual bootstrap}
\label{Lab: chain-ladder reserving and individual bootstrap}
\begin{tcolorbox}[title=]
  This section presents our two running examples (accident insurance and liability insurance) that will be revisited throughout the document. We compute their CL reserves, Mack's prediction uncertainty estimates, and we benchmark Mack's model error estimates by individual claims history bootstrap estimates, see Table \ref{CL results}.
\end{tcolorbox}

In this document, we study two small-scale examples. These small-scale examples provide a proof of concept, and generalization to bigger datasets still needs to be confirmed. To be able to perform a proper proof of concept, we select comparably old data such that not only the upper triangles are observed, but in these two datasets also the lower triangles are known. Thus, any method that we develop on the upper triangle can be benchmarked against the ground truth in the lower triangle in our two examples. Of course, this is very useful for providing evidence that our proposals work; generally, the results that require knowledge of the lower triangle are earmarked by an upper index $^\ddagger$ in this document, see, e.g., Table \ref{CL results}, below. We start by presenting the two datasets, these are the same as in Richman--W\"uthrich \cite{PtU}, and we also copy-paste the explaining text from that reference to describe the data.

\subsubsection{Accident insurance data}
The first dataset considers accident insurance on an annual scale with 5 fully observed accident years, i.e., we have a fully observed $5\times 5$ square. For model fitting and forecasting, we {\it only} use the {\it upper triangle},
as in Figure \ref{fig:CL2}, and we benchmark the forecasts against the true ultimates which are available here (having also observed the lower triangle).

\begin{table}[h]
\footnotesize
\centering
\begin{tabular}{lc}
\toprule
\textbf{Characteristic} &  \\
\midrule
Time scale & calendar years  \\
Number of accident years & 5  \\
Number of development years & 5  \\
Number of reported claims & 66,639  \\
\midrule
\multicolumn{2}{l}{\textbf{Data description}} \\
\midrule
\multicolumn{2}{l}{Annual individual cumulative payments $C_{i,j|\nu}$} \\
\multicolumn{2}{l}{Claim status $O_{i,j|\nu}\in \{0,1\}$ for closed/open at the end of period $j$} \\
\multicolumn{2}{l}{Binary static covariate for work or leisure accident} \\
\multicolumn{2}{l}{Calendar month of accident occurrence} \\
\multicolumn{2}{l}{Reporting delay in daily units} \\
\bottomrule
\end{tabular}
\caption{Characteristics of accident dataset.}
\label{tab:accidentdata}
\end{table}
Table \ref{tab:accidentdata} shows the available data. There are 66,639 reported claims with a fully observed development history over the $5\times 5$ square. Besides the individual cumulative payment process $C_{i,0:4|\nu}$, there is information about the claim status process $O_{i,0:4|\nu}$, with $O_{i,j|\nu}=1$ meaning that the $\nu$-th claim of accident year $i$ is open at the end of settlement delay $j$, and closed otherwise. Then, there is static information about: work or leisure related accident, the calendar month of the accident and the reporting delay in daily units. For more information we refer to Richman--W\"uthrich \cite{PtU}.

\subsubsection{Liability insurance data}
The second dataset considers liability insurance. We again have a fully observed $5\times 5$ square and for model fitting we only use the upper triangle.

\begin{table}[h]
\footnotesize
\centering
\begin{tabular}{lc}
\toprule
\textbf{Characteristic} &  \\
\midrule
Time scale & calendar years  \\
Number of accident years & 5  \\
Number of development years & 5  \\
Number of reported claims & 21,991  \\
\midrule
\multicolumn{2}{l}{\textbf{Data description}} \\
\midrule
\multicolumn{2}{l}{Annual individual cumulative payments $C_{i,j|\nu}$} \\
\multicolumn{2}{l}{Claim status $O_{i,j|\nu}\in \{0,1\}$ for closed/open  at the end of period $j$} \\
\multicolumn{2}{l}{Claims incurred $I_{i,j|\nu}\ge 0$} \\
\multicolumn{2}{l}{Binary static covariate for private vs.~commercial liability} \\
\multicolumn{2}{l}{Calendar month of accident occurrence} \\
\multicolumn{2}{l}{Reporting delay in daily units} \\
\bottomrule
\end{tabular}
\caption{Characteristics of liability dataset.}
\label{tab:liablitydata}
\end{table}

Table \ref{tab:liablitydata} shows the available data of the liability insurance dataset. The main difference to the previous example is that for this dataset there is also a claims incurred process $I_{i,0:4|\nu}$ available. The claims incurred process is a claims adjuster's prediction of the individual ultimate claim that is continuously updated when new information arrives, i.e., this is a stochastic process driven by the claims adjuster's assessments. 

\subsubsection{Mack's chain-ladder method and individual bootstrapping}
We start with Mack's \cite{Mack} distribution-free CL method. It allows one to compute the CL reserves at the evaluation date $I$ for each accident year $i > I-J$, given by
\begin{equation*}
\widehat{R}^{\rm CL}_{i} = \widehat{C}^{\rm CL}_{i,J} - C_{i,I-i}.
\end{equation*}
These CL reserves are benchmarked against the true {\it outstanding loss liabilities} (OLL), given by
${\rm OLL}_i=C_{i,J}- C_{i,I-i}$. These true OLL present the ground truth, and they are given in our small-scale examples because we know the lower triangles.
Table \ref{CL results} shows the CL results for the two datasets summed over all accident years $i$, and the column `Error$^\ddagger$' gives the total {\it forecast error}
\begin{equation}\label{forecast error}
\sum_{i=I-J+1}^I\widehat{C}^{\rm CL}_{i,J}-C_{i,J}.
\end{equation}

\begin{table}[h]
\centering
{\footnotesize
\begin{center}
\begin{tabular}{|l|r|r|rrr|rr|}
\hline
 & True OLL$^\ddagger$ & CL Reserves&Proc.Unc.&Est.Err.  & RMSEP & Error$^\ddagger$ &  \% RMSEP$^\ddagger$\\

\hline\hline
\underline{Accident dataset} &&&&&&&\\
Mack's CL model \cite{Mack} & 24,212&	23,064&1,429&851&1,663&-1,148 &69\%\\
Individual bootstrap & 24,212&	22,988&--&937&--&-1,224 &--\\
 \hline\hline
\underline{Liability dataset} &&&&&&&\\
Mack's CL model \cite{Mack} & 15,730&	11,526&1,383&1,413&	1,977&	-4,204&	213\%\\
Individual bootstrap & 15,730&	11,531&--&1,201&--&-4,199 &--\\
 \hline
\end{tabular}
\end{center}}
\caption{Mack's CL results on cumulative payments and CL results using an individual claims history bootstrap; the earmarked columns$^\ddagger$ can only be computed because we know the lower triangle in our two examples.}
\label{CL results}
\end{table}

We observe that in both datasets we underestimate the true OLL by -1,148 and -4,204, respectively, see column `Error$^\ddagger$' corresponding to \eqref{forecast error}. To assess the magnitude of this underestimation, we additionally compute Mack's \cite{Mack} rooted mean squared error of prediction (RMSEP), given by the square root of
the conditional MSEP 
\begin{eqnarray*}
\operatorname{msep}_{\sum_i C_{i,J}| {\cal D}_I}\left(\sum\nolimits_i
\widehat{C}^{\rm CL}_{i,J}\right)
&=& \E \left[\left.\left(\sum_{i=I-J+1}^I\widehat{C}^{\rm CL}_{i,J}-C_{i,J}
\right)^2\right| {\cal D}_I \right]
\\&=&
\underbrace{\operatorname{Var}\left(\left.\sum_{i=I-J+1}^IC_{i,J}\right| {\cal D}_I \right)}_{\text{\it process uncertainty}}
+\underbrace{\left(\sum_{i=I-J+1}^I\widehat{C}^{\rm CL}_{i,J}-\E \left.\left[C_{i,J}\right|{\cal D}_I\right]
\right)^2}_{\text{\it estimation error}},
\end{eqnarray*}
where ${\cal D}_I$ refers to the available cumulative payments at time $I$, see \eqref{available data at time I}. One of the main achievements of Mack \cite{Mack} was to compute/estimate the (rooted) {\it process uncertainty} (`Proc.Unc.'; also called {\it irreducible risk}) and the (rooted) {\it estimation error} (`Est.Err.'; also called {\it model error}) under suitable CL assumptions. This then provides the RMSEP. The numerical results are presented in columns `Proc.Unc.', `Est.Err.' and `RMSEP' of Table \ref{CL results} -- we always show the rooted versions. We observe that the 
forecast error \eqref{forecast error} makes 
$1,148/1,663=69\%$ of the RMSEP in the accident insurance case, this is a reasonable deviation (less than one RMSEP), and we cannot reject the CL method in this case.
In the liability insurance case, the CL method seems to perform worse, 
the forecast error \eqref{forecast error} makes 
$4,204/1,977=213\%$ of the RMSEP. This may lead us to doubt the application of the CL algorithm for the  liability insurance data.\footnote{The RMSEP is on the level of a standard deviation, so we typically check whether it exceeds two standard deviations (RMSEPs) or not.}

\medskip

Next, we present an individual claims history (non-parametric) bootstrap. In our context, a non-parametric bootstrap is useful to assess the (rooted) estimation error `Est.Err.', i.e., it is useful to analyze the model estimation uncertainty term by re-sampling new upper triangles to evaluate the resulting fluctuations in the CL factor estimates. It is not directly possible to assess the process uncertainty term with an individual claims history bootstrap in our set-up. The issue is that only the oldest accident periods $i\le I-J$ have {\it observed} ultimate claims $C_{i,J}$ (last column of upper triangle/trapezoid, see Figure \ref{Fig: individual claims RBNS}), and for all other accident periods $i=I-J+1,\ldots, I$ we cannot re-sample (bootstrap) ultimate claims. If the ultimate claim observations
$C_{i,J}$, $i\le I-J$, are sufficiently rich, we can project those to the more recent accident periods, otherwise we would not recommend the bootstrap to assess the process uncertainty term, but only the (rooted) estimation error `Est.Err.'. The (rooted) estimation error `Est.Err.' can be rewritten as
\begin{equation*}
\sum_{i=I-J+1}^I\widehat{C}^{\rm CL}_{i,J}-\E \left.\left[C_{i,J}\right|{\cal D}_I\right]
=  \sum_{i=I-J+1}^I C_{i,I-i} \left( \prod_{j=I-i}^{J-1}
\widehat{f}_j^{\rm CL} - \prod_{j=I-i}^{J-1}f_j\right).
\end{equation*}
For the non-parametric bootstrap, we randomly draw individual claims 
$C_{i,0:I-i|\nu}=(C_{i,j|\nu})_{j=0}^{I-i}$ with replacement from the upper individual claims triangle, see Figure \ref{Fig: individual claims RBNS}, such that the bootstrap sample has the same size as the original data sample of individual claims -- we perform this drawing with replacement simultaneously over all accident periods $i\in\{1,\ldots, I\}$ which also introduces some volatility across the accident periods. The resulting bootstrap sample is used to re-estimate the CL factors -- by first aggregating the individual bootstrapped claims similarly to \eqref{individual to aggregate} resulting in bootstrapped aggregated cumulative payments $C_{i,j}^\ast$, $i+j \le I$ -- and these are then used to compute the estimated bootstrapped CL factors $(\widehat{f}_j^{\ast})_{j=0}^{J-1}$ similar to \eqref{CL factor estimates}. Then, we compute the bootstrapped ultimate claim predictors by
\begin{equation}\label{bootstrap extrapolation}
\widehat{C}_{i,J}^\ast = C_{i,I-i} \prod_{j=I-i}^{J-1}\widehat{f}_j^{\ast},
\end{equation}
note that the basis $C_{i,I-i}$ remains fixed, as this corresponds to the conditioning on ${\cal D}_I$ in the RMSEP, i.e., we do not re-simulate the last diagonal of the upper triangle, we only bootstrap in order to assess the estimation uncertainty in the CL factor estimates. Repeating this re-estimation procedure many times, allows us to assess the average and the standard deviation in the bootstrap predictors $\widehat{C}_{i,J}^\ast$, the latter being an estimation uncertainty estimate. We report these bootstrap results (received from 1,000 bootstrap samples) on the lines `Individual bootstrap' in Table \ref{CL results}. The average bootstrap prediction is very well aligned with the original CL predictors $\widehat{C}^{\rm CL}_{i,J}$, thus, the bootstrap does not indicate any bias. The magnitude of the bootstrap standard deviation aligns well with the rooted estimation error estimate of Mack \cite{Mack}, we have a slightly higher value in the accident dataset (937 vs.~851) and a lower value in the liability dataset (1,201 vs.~1,413), but overall the magnitudes align.

In the above non-parametric bootstrap analysis, we resample the entire upper triangle by drawing with replacement. Another interesting analysis would be to re-sample only one selected accident period $i$, this would allow one to assess the impact of a single atypical accident period on the entire claims reserves.

\section{Chain-ladder RBNS reserving}
\label{Chain-ladder RBNS reserving}
In a preliminary step towards individual claims reserving, we separate RBNS from IBNR claims. A main motivation for this initial step is that there is individual claims information \eqref{covariates individual process} available for RBNS claims, and we try to optimally use this information to predict the individual ultimates $C_{i,J|\nu}$ of RBNS claims $\nu$, $i+T_{i|\nu}\le I$. This is not the case for IBNR claims (because they are not reported yet) and only a collective prediction is possible, e.g., based on exposure information.

\subsection{Chain-ladder RBNS prediction}
\begin{tcolorbox}[title=]
  This section modifies the recursive one-shot CL prediction algorithm, see Algorithm \ref{algorithm: recursive CL one-shot}, such that it only predicts RBNS claims. This is achieved by considering consistent claim cohorts in extrapolation; see Algorithm \ref{algorithm: RBNS CL one-shot}.
\end{tcolorbox}

The CL predictions $\widehat{C}^{\rm CL}_{i,J}$ cover both the RBNS and the IBNR claims. This comes from the fact that we do not consider the identical claims cohorts in the CL factor estimates, the second term in the numerator in \eqref{CL factors alternative 2} corresponds to IBNR claims at development delay $j$, see Figure \ref{Fig: individual claims RBNS}. It is straightforward to correct for this, and to only consider RBNS claims. We give the one-shot PtU factor version in Algorithm \ref{algorithm: RBNS CL one-shot}, as this is more convenient.

\medskip

\begin{algorithm}
\caption{Recursive one-shot CL RBNS prediction algorithm.}\label{algorithm: RBNS CL one-shot}
\begin{itemize}
\item[(a)]{\it Initialization for $j=J$.}
For the fully settled accident periods $i \in \{1,\ldots, I-J\}$, initialize the  algorithm by setting $\widehat{C}^{\rm RBNS}_{i,J|\nu}=C_{i,J|\nu}$ for all claims $\nu=1, \ldots, N_i$.
\item[(b)]{\it Iteration $j \to j-1 \ge 0$.} Compute recursively
\begin{equation}\label{PtU CL RBNS factor}
\widehat{F}^{\rm RBNS}_{j-1}= \frac{\sum_{i=1}^{I-j}\sum_{\nu:\, \mg{T_{i|\nu} \le j-1}}\widehat{C}^{\rm RBNS}_{i,J|\nu}}
{\sum_{i=1}^{I-j}\sum_{\nu:\, \mg{T_{i|\nu} \le j-1}}C_{i,j-1|\nu}}
\qquad \text{ and } \qquad \widehat{C}^{\rm RBNS}_{I-(j-1),J|\nu}=C_{I-(j-1),j-1|\nu}\,\widehat{F}^{\rm RBNS}_{j-1},
\end{equation}
for all RBNS claims $\nu$ at time $I$, i.e., with $T_{I-(j-1)|\nu} \le j-1$.
\end{itemize}
\end{algorithm}

Algorithm \ref{algorithm: RBNS CL one-shot} only extrapolates RBNS claims, and it does not add any margins for IBNR claims because the numerator and nominator of the PtU factors $\widehat{F}^{\rm RBNS}_{j-1}$ consider the {\it identical} RBNS claims cohort \mg{$\mg{T_{i|\nu} \le j-1}$}.

\subsection{Chain-ladder IBNR prediction}
\begin{tcolorbox}[title=]
  This section provides a partition of the total CL reserves into RBNS and IBNR reserves. This is  a natural consequence of the recursive one-shot CL RBNS prediction algorithm presented in the previous section in Algorithm \ref{algorithm: RBNS CL one-shot}.
\end{tcolorbox}

To forecast the IBNR claims, we can provide a similar algorithm. The IBNR reserves will consist of two different
terms in its estimation: (1) terms stemming from claims that are IBNR at time $I$, and (2) ultimate claims (estimates) that are used for IBNR prediction, but which are RBNS at time $I$. For this reason, the following paragraphs will use both upper indices $^{\rm IBNR}$ and $^{\rm RBNS}$.

\medskip

Initialize 
$\widehat{C}^{\rm RBNS}_{i,J|\nu}=C_{i,J|\nu}$ for all claims $\nu=1, \ldots, N_i$
in accident periods $i \in \{1, \ldots, I-J\}$ -- all these are RBNS claims at time $I$. The first recursive step
$J \to J-1$ considers
\begin{equation}\label{IBNR one-step}
\widehat{F}^{\rm IBNR}_{J-1}= \frac{\sum_{i=1}^{I-J}\sum_{\nu:\, \mg{T_{i|\nu} = J}}\widehat{C}^{\rm RBNS}_{i,J|\nu}}
{\sum_{i=1}^{I-J}\sum_{\nu:\, \mg{T_{i|\nu} \le J-1}}C_{i,J-1|\nu}}.
\end{equation}
This gives the (aggregated) IBNR claim prediction for accident period
$I-(J-1)$
\begin{equation*}
 \widehat{C}^{\mg{\rm IBNR}}_{I-(J-1),J}=\left[\sum_{\nu:\, T_{I-(J-1)|\nu} \le J-1} C_{I-(J-1),J-1|\nu}\right]\widehat{F}^{\rm IBNR}_{J-1}
 =C_{I-(J-1),J-1}\,\widehat{F}^{\rm IBNR}_{J-1}.
\end{equation*}
This considers the grossing-up factor from the observed cumulative payments $C_{I-(J-1),J-1}$ to the IBNR prediction for accident period
$I-(J-1)$. This can recursively be iterated, but the iteration is cumbersome. E.g., the next step 
$J-1 \to J-2$ looks as follows
\begin{equation*}
\widehat{F}^{\rm IBNR}_{J-2}= \frac{\sum_{i=1}^{I-(J-1)}\sum_{\nu:\, \mg{T_{i|\nu} = J-1}}\widehat{C}^{\rm RBNS}_{i,J|\nu}\,+\,\sum_{i=1}^{I-J}\sum_{\nu:\, \mg{T_{i|\nu} = J}}\widehat{C}^{\rm RBNS}_{i,J|\nu}\,+\,\widehat{C}^{\mg{\rm IBNR}}_{I-(J-1),J}}
{\sum_{i=1}^{I-(J-1)}\sum_{\nu:\, \mg{T_{i|\nu} \le J-2}}C_{i,J-2|\nu}}.
\end{equation*}
The first term in the numerator corresponds to the RBNS ultimate claim prediction of claims reported with delay $T_{i|\nu} = J-1$, the second term to the prediction of the claims reported with delay $T_{i|\nu} = J$ (all these claims are reported at time $I$), finally, the last term corresponds to the IBNR part corresponding to accident period $I-(J-1)$. Thus, we complete the upper-right triangle with IBNR predictions, and the missing part of this development rectangle is directly completed with the previous IBNR predictions $\widehat{C}^{\mg{\rm IBNR}}_{I-(J-1),J}$, so that all IBNR claims in the upper-right square/rectangle are identified.

A much easier way to receive the identical result is to subtract the RBNS ultimate claim predictors from the CL ones, that is, for all accident periods $i>I-J$ we have
\begin{equation}\label{IBNR computation}
 \widehat{C}^{\rm IBNR}_{i,J}
=  \widehat{C}^{\rm CL}_{i,J}-\sum_{\nu:\, T_{i|\nu} \le I-i}\widehat{C}^{\rm RBNS}_{i,J|\nu}.
\end{equation}
This gives a simple IBNR predictor for all accident periods.

\medskip

{\bf Remark.} A similar, though different approach is considered in Schnieper \cite{Schnieper}. The similarity concerns the fact that Schnieper \cite{Schnieper} also considers development ratios of type \eqref{IBNR one-step}, however, Schnieper \cite{Schnieper} uses an external exposure as nominator in \eqref{IBNR one-step}. Extrapolating this in a one-period ahead roll-over fashion is then peformed, this is doable but can be cumbersome (RBNS CL factors get contaminated by IBNR parts -- though in a mathematically consistent way). Having a past cumulative payments nominator and turning the problem to the one-shot ultimate prediction version allows us to receive an elegant decomposition \eqref{IBNR computation} in our set-up.

\subsection{Lab: Chain-ladder RBNS and IBNR reserving}
\begin{tcolorbox}[title=]
  This section provides the example how the CL reserves of Table \ref{CL results} can be partioned into RBNS reserves and IBNR reserves. This uses the RBNS Algorithm \ref{algorithm: RBNS CL one-shot} and the decomposition formula \eqref{IBNR computation}.
\end{tcolorbox}

We revisit the two examples introduced in Section \ref{Lab: chain-ladder reserving and individual bootstrap}. We apply Algorithm \ref{algorithm: RBNS CL one-shot} to compute the RBNS reserves. The IBNR reserves are then calculated as the differences \eqref{IBNR computation}. The results are presented in Table \ref{CL results RBNS and IBNR}.

\begin{table}[h]
\centering
{\footnotesize
\begin{center}
\begin{tabular}{|l|r|rr|rr|}
\hline
 & True OLL$^\ddagger$ & Reserves& RMSEP & Error$^\ddagger$ &  \% RMSEP$^\ddagger$\\
\hline\hline
\underline{Accident dataset} &&&&&\\
Mack's CL model \cite{Mack} & 24,212&	23,064&1,663&-1,148 &69\%\\
RBNS CL prediction, Algorithm \ref{algorithm: RBNS CL one-shot} & 19,735&	18,959&--&-774 &--\\
IBNR CL prediction \eqref{IBNR computation}& 4,478&	4,105&--&-374 &--\\
 \hline\hline
\underline{Liability dataset} &&&&&\\
Mack's CL model \cite{Mack} & 15,730&	11,526&	1,977&	-4,204&	213\%\\
RBNS CL prediction, Algorithm \ref{algorithm: RBNS CL one-shot} & 11,494&	8,601&--&-2,893 &--\\
IBNR CL prediction \eqref{IBNR computation}& 4,236&	2,925&--&-1,311 &--\\
 \hline
\end{tabular}
\end{center}}
\caption{Mack's CL results on cumulative payments split to RBNS and IBNR reserves; the earmarked columns$^\ddagger$ can only be computed because we know the lower triangle in our examples.}
\label{CL results RBNS and IBNR}
\end{table}

From the results in Table \ref{CL results RBNS and IBNR} we conclude that the CL method on RBNS claims seems to work very well for the accident insurance dataset. On the liability insurance dataset, the CL method seems to be negatively biased. We are going to refine this assessment in Section \ref{The role of claims incurred}, below. We use the `RBNS CL prediction' results of Table \ref{CL results RBNS and IBNR} as benchmarks for all subsequent individual claims reserving methods on RBNS claims.

%

\subsection{Individual claims reserving - setting the stage}

\begin{tcolorbox}[title=]
In theory, we are now fully prepared to dive into individual claims regression modeling. However, we still want to modify the weighted square loss minimization problem \eqref{weighted square loss} because in fine-grained regression problems, the solutions to the weighted square minimization may not be very robust in case $\max\{C_{i,j|\nu},\epsilon\}$ is small. This section introduces an unweighted square loss minimization problem, and we verify that the two problems give similar solutions (in the case without covariates). We do this in two steps, see Listings \ref{CLCode} and \ref{CLCode2}.
\end{tcolorbox}

{\bf Step 1.}
To set the stage for individual claims reserving, we first implement a modified version of Algorithm \ref{algorithm: RBNS CL one-shot}. Namely, we bring the PtU factor estimation in
\eqref{PtU CL RBNS factor} into a regression form so that it involves a weighted square loss minimization similar to \eqref{weighted square loss}.

\begin{lstlisting}[float=h,frame=tb,caption={Recursive one-shot CL RBNS algorithm with weighted square loss minimization \eqref{weighted square loss}.}, label=CLCode]
## initialize ultimate claims with observed ones for accident years i<=I0-J0
claims$YY <- NA
claims[which(claims$AccDate<=I0-J0),]$YY <- claims[which(claims$AccDate<=I0-J0),]$Ultimate

epsilon <- 0.001 ## avoid division by zero

## iterative PtU algorithm for RBNS prediction
for (j in rev(0:(J0-1))){
    i <- I0-j-1
    ### prepare learning data for GLM
    select <- which((claims$AccDate<=i)&(claims$RepDelayYY<=j))
    YY     <- claims[select,]$YY     ## has been filled in the previous loop
    CC     <- pmax(epsilon, triCC[select, paste0("X",j)])  ## these are cumulative cashflows
    learn  <- data.frame(cbind(YY, CC))
    ### prepare test data for PtU projection
    select <- which(claims$AccDate==(i+1))
    YY     <- claims[select,]$YY     ## is N/A
    CC     <- triCC[select, paste0("X",j)]
    test   <- data.frame(cbind(YY, CC))
    ### perform weighted square loss (G)LM - identity-link is used 
    glm1   <- glm(YY/CC ~ 1, start=0, weights=CC, data=learn, family=gaussian())
    claims[select, ]$YY <- test$CC * predict(glm1, newdata=test, type=c("response"))
       }
\end{lstlisting}

Listing \ref{CLCode} reformulates Algorithm \ref{algorithm: RBNS CL one-shot} with the PtU factor computation in \eqref{PtU CL RBNS factor} replaced by a linear regression (GLM with identity link) and using a weighted square loss minimization \eqref{weighted square loss} for model fitting; see line {\tt 21} of Listing \ref{CLCode}. Since RBNS claims can have individual cumulative payments $C_{i,j-1|\nu}$ being equal to zero, we selected a small positive constant $\epsilon=0.001$ to avoid dividing by zero, see lines {\tt 13} and {\tt 21} of Listing \ref{CLCode}. The solution of this algorithm gives the identical reserves as Algorithm \ref{algorithm: RBNS CL one-shot}, up to the
  $\epsilon>0$ correction factor, we also refer to \eqref{epsilon limit}.

\begin{table}[h]
\centering
{\footnotesize
\begin{center}
\begin{tabular}{|c|r|rr|r|r|}
\hline
$i$ & True OLL$^\ddagger$ & RBNS Algorithm \ref{algorithm: RBNS CL one-shot} & RBNS  Listing \ref{CLCode} &  Error$^\ddagger$ & Ind.RMSE$^\ddagger$  \\
\hline\hline
1 &0	&0	&0	&0	&0	\\
2&353&	339&	339&	-14	&	\bl{1.499}\\
3 &1,017&	1,305&	1,305&	288&\bl{2.956}\\
4 &3,102	&3,099	&3,099&-2&	\bl{4.263}\\
5 &15,263&	14,216	&14,216	&-1,046&	\bl{8.240}\\\hline
Total &19,735	&18,959	&18,959	&-774&\\		
 \hline
\end{tabular}
\end{center}}
\caption{Accident insurance: RBNS results of individual claims prediction using Algorithm \ref{algorithm: RBNS CL one-shot} and   Listing \ref{CLCode}; the earmarked columns$^\ddagger$ use the ground truth in the lower triangle.}
\label{CL results accident 1}
\end{table}

Table \ref{CL results accident 1} verifies that Algorithm \ref{algorithm: RBNS CL one-shot} and Listing \ref{CLCode} give the same results. The first column shows the true OLL for each accident years $i \in \{1,\ldots, 5\}$. Columns 
`RBNS Algorithm \ref{algorithm: RBNS CL one-shot}' and `RBNS  Listing \ref{CLCode}' verify that the two algorithms give the same results.
The column `Error$^\ddagger$' shows how the total RBNS forecast error of -774 splits across the different accident years $i\in \{1,\ldots, 5\}$, see also second line of Table \ref{CL results RBNS and IBNR}.

The final column `Ind.RMSE$^\ddagger$' of Table \ref{CL results accident 1} will be the quantity of major interest for all subsequent models. It considers the rooted mean square error (RMSE) on an {\it individual claims level}, that is, we define the {\it individual average RBNS prediction errors} by 
\begin{equation}\label{individual RMSE}
\text{Ind.RMSE}^\ddagger_i=
\sqrt{\frac{1}{\sum_{\nu:\, T_{i|\nu} \le I-i} 1} ~
\sum_{\nu:\, T_{i|\nu} \le I-i} \left(\widehat{C}^{\rm RBNS}_{i,J|\nu} - C_{i,J|\nu} \right)^2} \qquad \text{ for $i \ge I-J$.}
\end{equation}
This is the average prediction accuracy on the individual claims level (measured by the RMSE). Typically, with improved models, we expect these numbers to decrease. Remark that we can compute \eqref{individual RMSE} in our examples because we know the lower triangle, earmarked by $^\ddagger$.

\bigskip

{\bf Step 2.}
Before starting with individual claims reserving, we still modify the algorithm once more. Namely, we want to remove the weighting in the square loss minimization to robustify the prediction algorithm (this also allows us to get rid of the constant $\epsilon>0$). For this, we replace the weighted square minimization \eqref{weighted square loss} by the following {\it linear regression problem} for RBNS reserving
\begin{equation}\label{weighted square loss 2}
\widehat{\bvartheta}_{j-1} =
\underset{\bvartheta=(\vartheta_0, \vartheta_1)^\top \in \R^2} {\arg\min} \left\{
\sum_{i=1}^{I-j}\sum_{\nu:\, \mg{T_{i|\nu} \le j-1}}
\left(\widehat{C}_{i,J|\nu}- \left(\vartheta_0 + \vartheta_1 C_{i,j-1|\nu}\right) \right)^2 \right\}.
\end{equation}
We add an intercept $\vartheta_0 \in \R$ and drop the weighting (for more robustness).
Naturally, this gives a different solution. We verify in Table \ref{CL results accident 1B} that the solution is close to the weighted version. Moreover, using the identity link in the square loss minimization \eqref{weighted square loss 2} 
implies that the (in-sample) balance property is fulfilled; see Lindholm--W\"uthrich \cite{LindholmW}. This is an important property that ensures bias control in our recursive estimation procedure.
The code is provided in Listing \ref{CLCode2}; and this is the basic code to dive into regression modeling for RBNS reserving.

\begin{lstlisting}[float=h,frame=tb,caption={Recursive one-shot RBNS reserving algorithm using a Gaussian linear regression \eqref{weighted square loss 2}.}, label=CLCode2]
## initialize ultimate claims with observed ones for accident years i<=I0-J0
claims$YY <- NA
claims[which(claims$AccDate<=I0-J0),]$YY <- claims[which(claims$AccDate<=I0-J0),]$Ultimate

## iterative PtU algorithm for RBNS prediction
for (j in rev(0:(J0-1))){
    i <- I0-j-1
    ### prepare learning data for GLM
    select <- which((claims$AccDate<=i)&(claims$RepDelayYY<=j))
    YY     <- claims[select,]$YY     ## has been filled in the previous loop
    CC     <- triCC[select, paste0("X",j)]  ## these are cumulative cashflows
    learn  <- data.frame(cbind(YY, CC))
    ### prepare test data for PtU projection
    select <- which(claims$AccDate==(i+1))
    YY     <- claims[select,]$YY     ## is N/A
    CC     <- triCC[select, paste0("X",j)]
    test   <- data.frame(cbind(YY, CC))
    ### perform square loss (G)LM - identity-link is used 
    glm2   <- glm(YY ~ CC, data=learn, family=gaussian())
    claims[select, ]$YY <- predict(glm2, newdata=test, type=c("response"))
       }
\end{lstlisting}

\begin{table}[h]
\centering
{\footnotesize
\begin{center}
\begin{tabular}{|c|r|rr|rr|rr|}
\hline
 &  & RBNS  & RBNS   &  Error$^\ddagger$ &Error$^\ddagger$ & Ind.RMSE$^\ddagger$& Ind.RMSE$^\ddagger$  \\
$i$ & True OLL$^\ddagger$ & Listing \ref{CLCode} & Listing \ref{CLCode2} &  Listing \ref{CLCode} & Listing \ref{CLCode2}&  Listing \ref{CLCode} & Listing \ref{CLCode2}  \\
\hline\hline
1 &0	&0	&0	&0	&0&0&0	\\
2&353&	339&	337&	-14	&	-16	&	1.499&	\bl{1.489}\\
3 &1,017&	1,305&	1,338&	288&321&2.956&\bl{2.985}\\
4 &3,102	&3,099	&3,264&-2&163&	4.263&	\bl{4.262}\\
5 &15,263&	14,216	&14,137	&-1,046&-1,126&	8.240&	\bl{8.218}\\\hline
Total &19,735	&18,959	&19,076	&-774&-658&&\\		
 \hline
\end{tabular}
\end{center}}
\caption{Accident insurance: RBNS results of individual claims prediction using  Listings \ref{CLCode} and \ref{CLCode2}; the earmarked columns$^\ddagger$ use the ground truth in the lower triangle.}
\label{CL results accident 1B}
\end{table}

Table \ref{CL results accident 1B} compares the results of the algorithms given in Listings \ref{CLCode} and \ref{CLCode2}. This verifies that the two algorithms provide very similar results. We have a preference for the second algorithm, as it is more robust and easy to extend. In fact, this similarity between the results of Listings \ref{CLCode} and \ref{CLCode2} should be checked case by case, and on small portfolios with volatile claims it might be violated.

\medskip

\begin{tcolorbox}[title=]
\begin{center}
We are now ready: The results of Table \ref{CL results accident 1B} serve as benchmark for all subsequent derivations on individual RBNS claims reserving (involving past claims histories).
\end{center}
\end{tcolorbox}

%
%

  
\section{Individual ultimate prediction - one-shot micro reserving}
\label{Individual ultimate prediction using machine learning}
We now turn our attention to ML applications for individual RBNS claims reserving.
For this we recall the individual claim settlement process \eqref{definition claims settlement process}, which is
given by
\begin{equation}
{\cal C}_{i|\nu}=\left[
\begin{pmatrix}C_{i,0|\nu}\\\bX_{i,0|\nu}\end{pmatrix}\mathds{1}_{\{T_{i|\nu} \le 0\}}
,\,
\begin{pmatrix}C_{i,1|\nu}\\\bX_{i,1|\nu}\end{pmatrix}\mathds{1}_{\{T_{i|\nu} \le 1\}}
,\ldots, \,
\begin{pmatrix}C_{i,J|\nu}\\\bX_{i,J|\nu}\end{pmatrix}\mathds{1}_{\{T_{i|\nu} \le J\}}
\right],
\end{equation}
all IBNR periods $j< T_{i|\nu}$ are masked, and the periods $i+j>I$ have not been observed yet, because they lie in the future at the evaluation date $I$.

\medskip

{\bf Assumption.} We assume that the individual claims processes ${\cal C}_{i|\nu}$ are independent, and that they are conditionally i.i.d., given the static covariates. 

\subsection{Recursive individual RBNS claims reserving}
\label{Recursive individual RBNS claims reserving}

\begin{tcolorbox}[title=]
  This section introduces the generic algorithm for recursive one-shot forecasting using general ML regression models.
  An important point is the consistent consideration of past information for estimating and forecasting RBNS claims, see
  learning sample \eqref{appended history}.
\end{tcolorbox}

In view of Algorithm \ref{algorithm: RBNS CL one-shot}
and Listing \ref{CLCode2}, it is obvious how to lift these algorithms to general ML regression models for RBNS reserving. Algorithm \ref{algorithm: PtU one-shot} gives the generic algorithm. The important point is that one always considers consistent cohorts for PtU factor estimation and projection.
This is indicated by the choice of the learning sample ${\cal L}_{j-1}$
in Algorithm \ref{algorithm: PtU one-shot}, see \eqref{appended history}, constraining the inputs by \mg{$T_{i|\nu} \le j-1$}.

\begin{algorithm}
\caption{Generic recursive one-shot PtU forecast algorithm for RBNS claims.}\label{algorithm: PtU one-shot}
\begin{itemize}
\item[(a)]{\it Initialization for $j=J$.}
For the fully settled accident periods $i \in \{1,\ldots, I-J\}$, initialize the  algorithm by setting $\widehat{C}_{i,J|\nu}=C_{i,J|\nu}$ for all claims $\nu=1, \ldots, N_i$.
\item[(b)]{\it Iteration $j \to j-1 \ge 0$.} 
\begin{itemize}
\item[(b1)] Select the learning sample
\begin{equation}
{\cal L}_{j-1} =
\left\{ \left(\widehat{C}_{i,J|\nu}, (C_{i,l|\nu}, \bX_{i,l|\nu})_{l=0}^{j-1}\right);\,
\mg{T_{i|\nu} \le j-1}
\text{ and } i\le I-j  \right\}.
\label{appended history}
\end{equation}
\item[(b2)] Fit a regression model $\mu_{j-1}$ on the learning sample ${\cal L}_{j-1}$ by using
\begin{equation}
(C_{i,l|\nu}, \bX_{i,l|\nu})_{l=0}^{j-1}
~\mapsto~ \mu_{j-1}\left((C_{i,l|\nu}, \bX_{i,l|\nu})_{l=0}^{j-1}\right)
=
\E \left[\widehat{C}_{i,J|\nu} \left|(C_{i,l|\nu}, \bX_{i,l|\nu})_{l=0}^{j-1}
\right]\right. .
 \label{estimate 2}
\end{equation}
\item[(b3)]
Compute the predictions of the RBNS claims $\nu$ of accident year $I-(j-1)$ by
\begin{equation}\label{prediction model PtU} 
\widehat{C}_{I-(j-1),J|\nu}=\mu_{j-1}\left((C_{I-(j-1),l|\nu}, \bX_{I-(j-1),l|\nu})_{l=0}^{j-1}\right).
\end{equation}
\end{itemize}
\end{itemize}
\end{algorithm}

\medskip

\begin{rems}[Algorithm \ref{algorithm: PtU one-shot}]\normalfont
\begin{itemize}
\item Listing \ref{CLCode2} is a special case of Algorithm \ref{algorithm: PtU one-shot}, where the only input information used is the latest individual cumulative payment, see \eqref{weighted square loss 2}. That is,
\begin{equation}\label{CL representation general}
\mu_{j-1}\left((C_{i,l|\nu}, \bX_{i,l|\nu})_{l=0}^{j-1}\right)=\mu_{j-1}\left(C_{i,j-1|\nu}\right).
\end{equation}
This makes it obvious how to lift Listing \ref{CLCode2} to a general ML forecast algorithm.
\item
Algorithm \ref{algorithm: PtU one-shot} describes one-shot PtU forecasting of RBNS claims. Recursively going from settlement period $j$ to period $j-1$, we aim at forecasting the RBNS claims of
accident periods $I-(j-1)$, see \eqref{prediction model PtU}. Since these claims are RBNS claims at time $I$, they can have a maximal reporting delay of
\mg{$T_{i|\nu} \le j-1$}. This is then reflected in the learning sample ${\cal L}_{j-1}$ by setting the corresponding side constraint, see \eqref{appended history}. Thus, the learning sample and the forecast problem consider the same side constraint in building their claims cohorts.
\item Since we do not know the true ultimate claims $C_{i,J|\nu}$ for accident periods $i>I-J$, we recursively replace them by their
forecasts $\widehat{C}_{i,J|\nu}$, see \eqref{appended history}. This is completely analogous to the one-shot RBNS prediction \eqref{PtU CL RBNS factor}. 
Using the tower property for conditional expectation, this is justified as follows
\begin{eqnarray*}
(C_{i,l|\nu}, \bX_{i,l|\nu})_{l=0}^{j-1}
~\mapsto~ \mu_{j-1}\left((C_{i,l|\nu}, \bX_{i,l|\nu})_{l=0}^{j-1}\right)
&=&\nonumber
\E \left[C_{i,J|\nu} \left|(C_{i,l|\nu}, \bX_{i,l|\nu})_{l=0}^{j-1}
\right]\right.
\\&=&
\E \left[\widehat{C}_{i,J|\nu} \left|(C_{i,l|\nu}, \bX_{i,l|\nu})_{l=0}^{j-1}
\right]\right. .
\end{eqnarray*}
The latter can be learned from the learning sample ${\cal L}_{j-1}$ given in \eqref{appended history}.
Recursive iteration completes the forecasting, and it is aligned with Figure \ref{fig:CL2}.
\item From Algorithm \ref{algorithm: PtU one-shot} it is obvious that this forecast procedure can deal with any (dynamic) input information, in particular, we can also consider continuous time inputs  \eqref{continuous time input} and unstructured data.
\item In practical applications, the only critical item of this algorithm is its recursive nature. In particular, we need to perform a careful bias control because a bias can easily propagate through the recursive forecast architecture. 
In Listing \ref{CLCode2} we consider a Gaussian linear regression problem, and MLE provides the in-sample balance property, thus, it provides an in-sample guarantee of unbiasedness. For more complex ML algorithms we need to enforce this balance property manually, e.g., by a post correction or by regularization, we come back to this in \eqref{post calibration}, below.
\end{itemize}
\end{rems}

\subsection{Lab: Accident insurance example -- linear regression}
\label{Lab: Accident insurance example -- linear regression}
\begin{tcolorbox}[title=]
This section gives first explicit examples of the one-shot PtU forecast Algorithm \ref{algorithm: PtU one-shot}. It uses a (simple) linear rgression on the available covariates of the last observed period. The first example in Table \ref{CL results accident 2A} only considers individual cumulative payments and the claim status, the second example in Table \ref{CL results accident 2B} considers all available covariates at time $j-1$.
\end{tcolorbox}

We revisit the example of Table \ref{CL results accident 1B} and we challenge the results by more complex regression models based on Algorithm
\ref{algorithm: PtU one-shot}.

\begin{lstlisting}[float=h,frame=tb,caption={Recursive one-shot PtU RBNS algorithm  including the latest claim status.}, label=CLCode3]
## initialize ultimate claims with observed ones for accident years i<=I0-J0
claims$YY <- NA
claims[which(claims$AccDate<=I0-J0),]$YY <- claims[which(claims$AccDate<=I0-J0),]$Ultimate

## iterative PtU algorithm for RBNS prediction
for (j in rev(0:(J0-1))){
    i <- I0-j-1
    ### prepare learning data for GLM
    select <- which((claims$AccDate<=i)&(claims$RepDelayYY<=j))
    YY     <- claims[select,]$YY     ## has been filled in the previous loop
    CC     <- triCC[select, paste0("X",j)]  ## these are cumulative cashflows
    Status <- triOO[select, paste0("X",j)]  ## claim status process
    learn  <- data.frame(cbind(YY, CC, Status))
    ### prepare test data for PtU projection
    select <- which(claims$AccDate==(i+1))
    YY     <- claims[select,]$YY     ## is N/A
    CC     <- triCC[select, paste0("X",j)]
    Status <- triOO[select, paste0("X",j)]
    test   <- data.frame(cbind(YY, CC, Status))
    ### perform square loss (G)LM - identity-link is used 
    glm2   <- glm(YY ~ CC*Status, data=learn, family=gaussian())
    claims[select, ]$YY <- predict(glm2, newdata=test, type=c("response"))
       }
\end{lstlisting}

We start by a simple linear regression model that only additionally considers the latest claim status 
$O_{i,j-1|\nu}\in \{0,1\}$ in the input information. Thus, we consider whether the $\nu$-th claim of accident period $i$ is closed or open after settlement delay $j-1$.
The reason for this choice is that the claim status information is the most important one to forecast whether there are more payments on a given claim.
For the regression function \eqref{estimate 2}, we select a simple linear regression model with an interaction term, that is, we set
\begin{eqnarray}\nonumber
\mu_{j-1}\left((C_{i,l|\nu}, \bX_{i,l|\nu})_{l=0}^{j-1}\right)
&=& \vartheta_0 + \vartheta_1 \,C_{i,j-1|\nu}+ \vartheta_2 \,O_{i,j-1|\nu}+ \vartheta_3\, C_{i,j-1|\nu}\,O_{i,j-1|\nu}
\\&=&\label{linear regression claim status}
\left(\vartheta_0 + \vartheta_2 \,O_{i,j-1|\nu}\right) + \left(\vartheta_1 + \vartheta_3\, O_{i,j-1|\nu}\right)C_{i,j-1|\nu},
\end{eqnarray}
for regression parameter $(\vartheta_k)_{k=0}^3 \in \R^4$.
Basically, this means that open claims are regressed with parameters $\vartheta_0+\vartheta_2$ and $\vartheta_1+\vartheta_3$, and closed claims are regressed with parameters $\vartheta_0$ and $\vartheta_1$. This is implemented in Listing \ref{CLCode3}, and the results are shown in Table \ref{CL results accident 2A}.

\begin{table}[h]
\centering
{\footnotesize
\begin{center}
\begin{tabular}{|c|r|rr|rr|rr|}
\hline
 &  & RBNS  & RBNS   &  Error$^\ddagger$ &Error$^\ddagger$ & Ind.RMSE$^\ddagger$& Ind.RMSE$^\ddagger$  \\
$i$ & True OLL$^\ddagger$ & Listing \ref{CLCode} & Listing \ref{CLCode3} &  Listing \ref{CLCode} & Listing \ref{CLCode3}&  Listing \ref{CLCode} & Listing \ref{CLCode3}  \\
\hline\hline
1 &0	&0	&0	&0	&0&0&0	\\
2&353&	339&	388&	-14	&	36	&	1.499&	\bl{1.455}\\
3 &1,017&	1,305&	1,407&	288&390&2.956&\bl{3.012}\\
4 &3,102	&3,099	&3,285&-2&183&	4.263&	\bl{4.221}\\
5 &15,263&	14,216	&15,000	&-1,046&-263&	8.240&\bl{8.135}\\\hline
Total &19,735	&18,959	&20,080	&-774&346&&\\		
 \hline
\end{tabular}
\end{center}}
\caption{Accident insurance: RBNS results of individual claims prediction using  Listings \ref{CLCode} and \ref{CLCode3}, the latter adds a linear regression on the latest claim status information $O_{i,j-1|\nu} \in \{0,1\}$, see \eqref{linear regression claim status}; the earmarked columns$^\ddagger$ use the ground truth
in the lower triangle.}
\label{CL results accident 2A}
\end{table}

We observe a significant improvement of the individual claim RMSEs
(column `Ind.RMSE$^\ddagger$', in \bl{blue color})  except in accident period $i=3$. This shows that the latest claim status is important information to forecast further payments on a given claim $\nu$. Interestingly, these results (with identity link) outperform the neural network results (with log-link) of Richman--W\"uthrich \cite[Table 6]{PtU}. This shows that the network results in that reference can be improved. Our experiments have shown that the identity link leads to better results than the log-link in this accident insurance data example. The identity link does not guarantee non-negativity of ultimate claims, the log-link does not allow ultimate claims to be exactly equal to zero. Thus, both choices have deficiencies and it remains an open problem to improve on this point.

\begin{figure}[htb!]
\begin{center}
\begin{minipage}[t]{0.49\textwidth}
\begin{center}
\includegraphics[width=\textwidth]{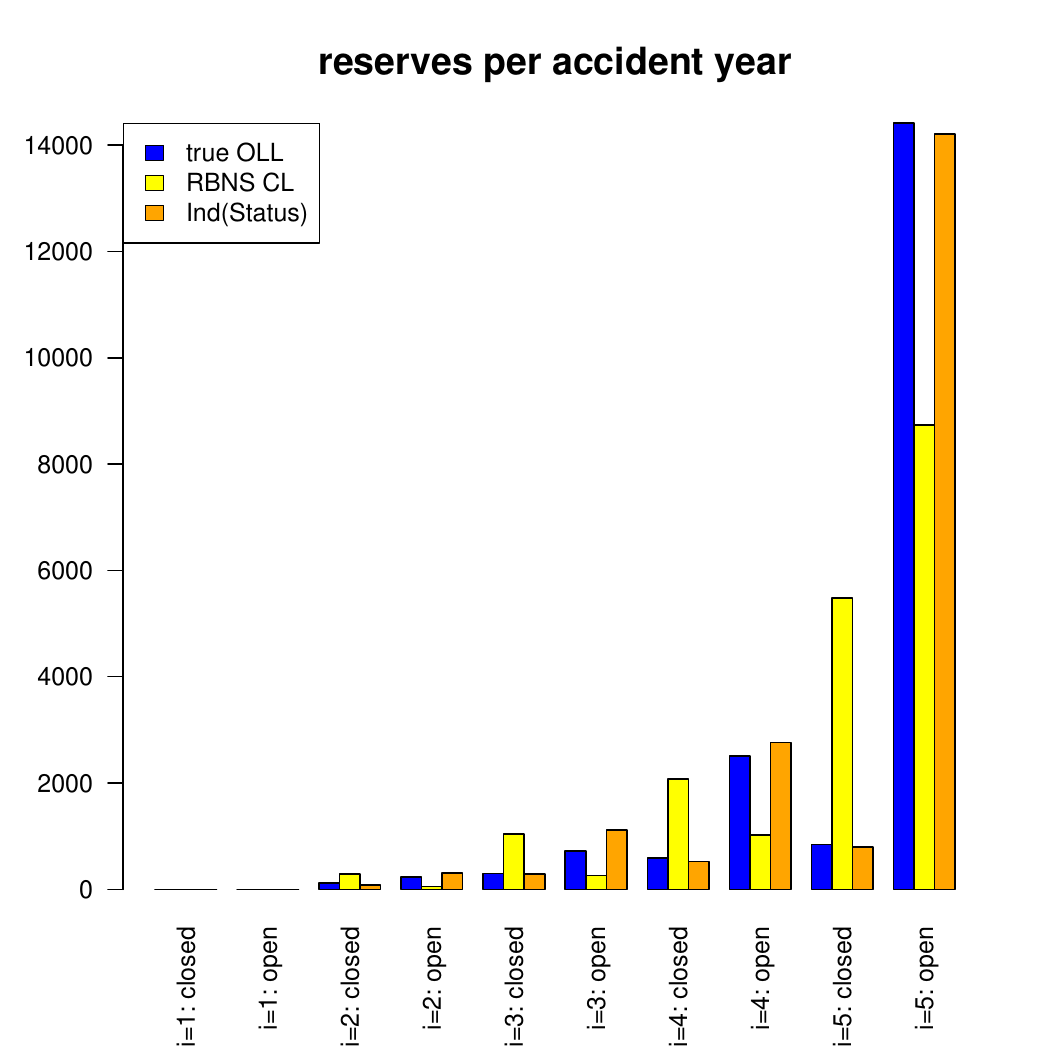}
\end{center}
\end{minipage}
\end{center}
\caption{Claims reserves per accident year $i=1,\ldots, 5$ and separated by closed and open claims at the evaluation date $I$ using the linear regression \eqref{linear regression claim status}. }
\label{figure reserves open claims}
\end{figure}

Figure \ref{figure reserves open claims} illustrates the results
of Listing \ref{CLCode3}. It shows the resulting claims reserves per accident year $i=1,\ldots, 5$ and split according to the claim status $O_{i,I-i|\nu} \in \{0,1\}$ (closed/open) at the evaluation date $I$. We observe that the claims reserves of 
Listing \ref{CLCode3} (in orange color) meet the true OLL (in blue color) very well (this is an out-of-sample consideration, evaluating on the ground truth OLL), whereas the CL  RBNS method (in yellow color) cannot distinguish between closed and open claims.
This verifies the improvements reported in Table \ref{CL results accident 2A}.

\medskip

We extend the regression model given in \eqref{linear regression claim status} to include all available information (covariates) of period $j-1$. That is, we make a Markov assumption and the latest information is again included in a linear regression model (with identity link)
\begin{eqnarray}\label{linear regression number 2}
\mu_{j-1}\left((C_{i,l|\nu}, \bX_{i,l|\nu})_{l=0}^{j-1}\right)
&=& \mu_{j-1}\left(C_{i,j-1|\nu}, \bX_{i,j-1|\nu}\right)
\\&=& \vartheta_0 + \vartheta_1 \,C_{i,j-1|\nu}+ \vartheta_2 \,C_{i,j-1|\nu}\,O_{i,j-1|\nu}
+ \sum_{k \ge 1} \vartheta_{k+2} \, X^{(k)}_{i,j-1|\nu},
\nonumber
\end{eqnarray}
the last sum considers the components of $\bX_{i,j-1|\nu}$ as a linear regression, 
we use dummy coding for the calendar month of the accident date, and the reporting delay is censored at 365 days; see Table \ref{tab:accidentdata} for the available covariates.
Moreover, we keep the interaction term between the individual cumulative payments $C_{i,j-1|\nu}$ and the claim status $O_{i,j-1|\nu}$. The results are reported in Table \ref{CL results accident 2B}, and the fitting results of the linear regression model of the last period $j-1=0$ (i.e., $i=5$) are shown in Listing \ref{GLMOutput}; we comment on this below.

\begin{table}[h]
\centering
{\footnotesize
\begin{center}
\begin{tabular}{|c|r|rr|rr|rr|}
\hline
 &  & RBNS  & RBNS   &  Error$^\ddagger$ &Error$^\ddagger$ & Ind.RMSE$^\ddagger$& Ind.RMSE$^\ddagger$  \\
$i$ & True OLL$^\ddagger$ & Listing \ref{CLCode} & All covariates &  Listing \ref{CLCode} & All covariates&  Listing \ref{CLCode} & All covariates  \\
\hline\hline
1 &0	&0	&0	&0	&0&0&0	\\
2&353&	339&	374&	-14	&	22	&	1.499&	\bl{1.455}\\
3 &1,017&	1,305&	1,411&	288&394&2.956&\bl{3.013}\\
4 &3,102	&3,099	&3,358&-2&256&	4.263&	\bl{4.221}\\
5 &15,263&	14,216	&14,965	&-1,046&-298&	8.240&\bl{8.121}\\\hline
Total &19,735	&18,959	&20,108	&-774&374&&\\		
 \hline
\end{tabular}
\end{center}}
\caption{Accident insurance: RBNS results of individual claims prediction using  Listing \ref{CLCode} and the linear regression \eqref{linear regression number 2} on all available covariates of settlement period $j-1$;
 the earmarked columns$^\ddagger$ use the ground truth
in the lower triangle.}
\label{CL results accident 2B}
\end{table}

Comparing Tables \ref{CL results accident 2A} and \ref{CL results accident 2B}, we observe a huge similarity between the results, there is only one improvement in 
`Ind.RMSE$^\ddagger$' for the most recent accident year $i=5$ (compare \bl{blue colors} in both tables). This verifies that the individual cumulative payments and the claim status are the most important covariates in this forecast, and the remaining covariates give some further fine-tuning for the most recent accident year. Listing \ref{GLMOutput} shows the regression output of the last linear regression function $\mu_0$, i.e., for $j=0$, which is used to extrapolate the most recent accident year $i=5$. From a quick inspection we conclude that we may drop the input variable 'work or leisure accident', and all the other variables should remain in the linear regression model.

\begin{lstlisting}[float=h,frame=tb,caption={GLM output of the linear regression function \eqref{linear regression number 2} for $\mu_{0}$ (i.e., $j=0$, resp., $i=5$).}, label=GLMOutput]
Call:
glm(formula = YY ~ CC * Status + WorkLeisure + AccMonth + 
    RepDelay, family = gaussian(), data = learn)

Coefficients:
                Estimate Std. Error t value Pr(>!t!)    
(Intercept)   -0.2198331  0.1704315  -1.290  0.19711    
CC             1.0672223  0.0131790  80.979  < 2e-16 ***
Status         0.6921395  0.0695684   9.949  < 2e-16 ***
WorkLeisure   -0.0242436  0.0223238  -1.086  0.27749    
AccMonth2      0.0146845  0.1108452   0.132  0.89461    
AccMonth3      0.0398143  0.1136975   0.350  0.72621    
AccMonth4      0.1649879  0.1205377   1.369  0.17108    
AccMonth5      0.2224703  0.1162873   1.913  0.05574 .  
AccMonth6      0.0850967  0.1166250   0.730  0.46560    
AccMonth7      0.3449664  0.1196346   2.883  0.00393 ** 
AccMonth8      0.7057549  0.1179919   5.981 2.23e-09 ***
AccMonth9      0.8631275  0.1262528   6.837 8.22e-12 ***
AccMonth10     1.1990894  0.1348060   8.895  < 2e-16 ***
AccMonth11     1.8606903  0.1469851  12.659  < 2e-16 ***
AccMonth12     2.0240857  0.2085132   9.707  < 2e-16 ***
RepDelay       0.0045515  0.0007148   6.367 1.94e-10 ***
CC:Status      0.4165162  0.0148093  28.125  < 2e-16 ***
---
Signif. codes:  0 '***' 0.001 '**' 0.01 '*' 0.05 '.' 0.1 ' ' 1

(Dispersion parameter for gaussian family taken to be 30.99334)

    Null deviance: 3141607  on 45898  degrees of freedom
Residual deviance: 1422036  on 45882  degrees of freedom
AIC: 287881

Number of Fisher Scoring iterations: 2
\end{lstlisting}

From Listing \ref{GLMOutput} we observe that
the OLL prediction is increasing in the 'accident month' variable. This makes sense for the most recent accident year, as accident month 'January' has a 12-months development by the end of the calendar year, and accident month 'December' only a 1-month development. So, we expected more open payments for later accidents during the calendar year (because they are less developed caused by the accounting year cut-off). The variable 'reporting delay' also leads to increasing claims, this may be caused by the fact that longer reporting delays correlate with longer waiting periods, and hence larger claims (because they are more severe).

\begin{figure}[htb!]
\begin{center}
\begin{minipage}[t]{0.49\textwidth}
\begin{center}
\includegraphics[width=\textwidth]{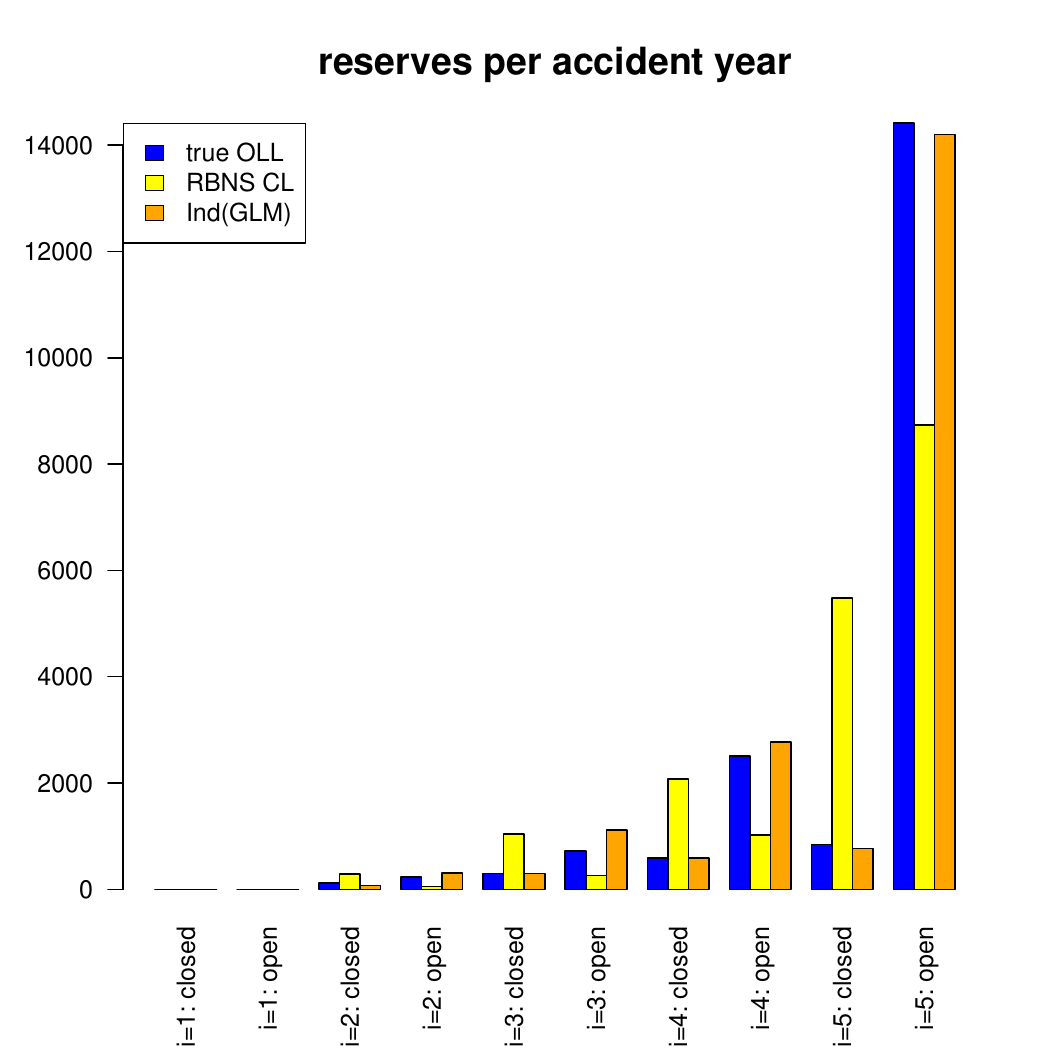}
\end{center}
\end{minipage}
\begin{minipage}[t]{0.49\textwidth}
\begin{center}
\includegraphics[width=\textwidth]{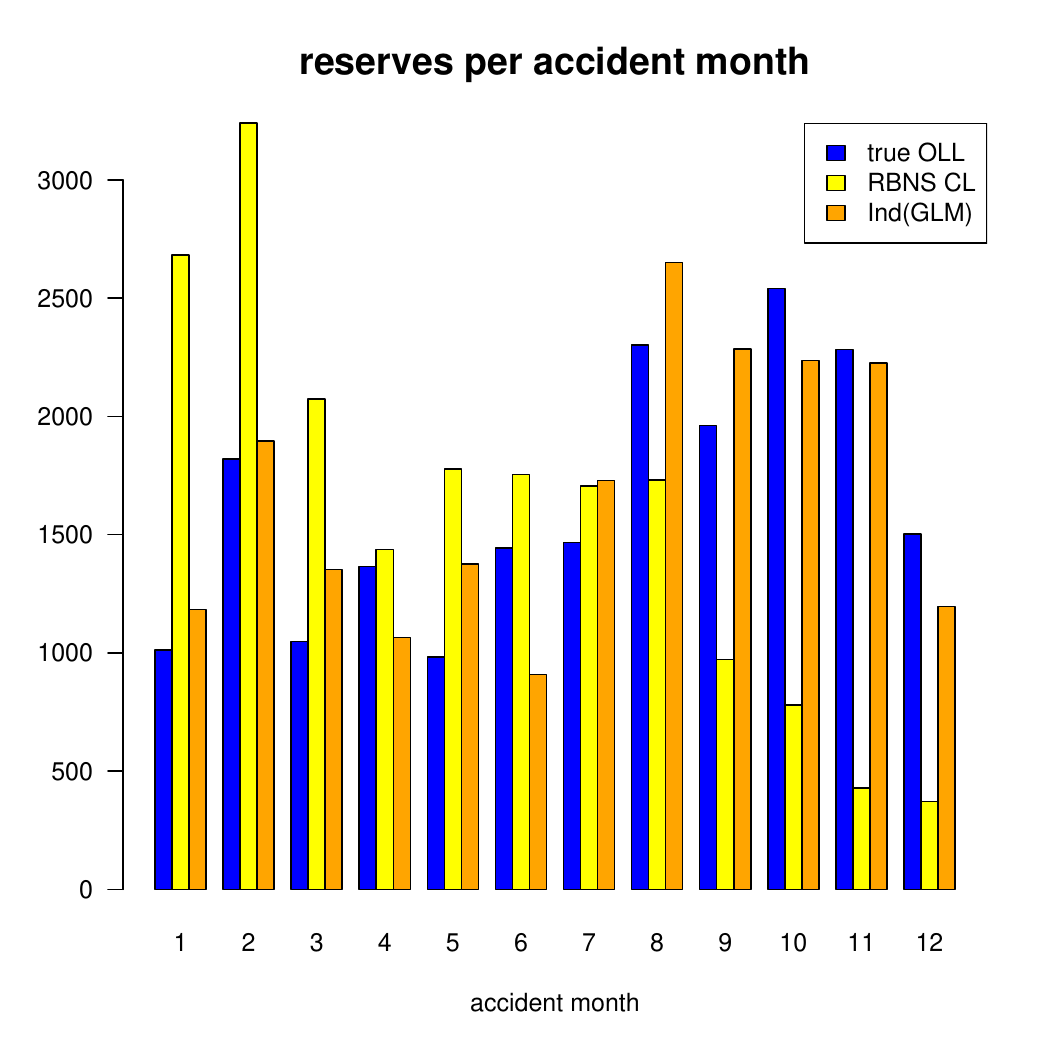}
\end{center}
\end{minipage}
\end{center}
\caption{(lhs) Claims reserves per accident year $i=1,\ldots, 5$ and separated by closed and open claims at the evaluation date $I$, (rhs) claims reserves split w.r.t.~the accident month both graphs using the linear regression \eqref{linear regression number 2}, see also Listing \ref{GLMOutput}. }
\label{figure reserves open claims 2}
\end{figure}

Figure \ref{figure reserves open claims 2} shows the resulting claims reserves of the linear regression model \eqref{linear regression number 2}, see also Listing \ref{GLMOutput}. The figure on the left-hand side splits the reserves w.r.t.~the accident years and the claim status, and on the right-hand side the reserves are split w.r.t.~the accident month. We observe that the estimated reserves (in orange color) are well aligned with the true outcomes (in blue color) saying that we have rather accurate forecasts on the different covariate levels. On the other hand, the RBNS CL reserves (in yellow color) cannot cope with this behavior.

\subsection{Lab: Linear regression bootstrap results}
\label{Lab: Linear regression bootstrap results}
\begin{tcolorbox}[title=]
Since the linear regressions of Tables \ref{CL results accident 2A} and \ref{CL results accident 2B} can be computed very fast, this allows us to run an individual claims history bootstrap to analyze model estimation uncertainty. This section presents the bootstrap results for the linear regression case.
\end{tcolorbox}

The individual claims reserving results of Tables \ref{CL results accident 1B}, \ref{CL results accident 2A} and \ref{CL results accident 2B} can be computed very fast -- each one involves only 4 linear regressions. This makes it feasible to run an individual claims bootstrap analysis, similar to Section \ref{Lab: chain-ladder reserving and individual bootstrap}.

\begin{table}[h]
\centering
{\footnotesize
\begin{center}
\begin{tabular}{|l|r|rr|rr|}
\hline
&&\multicolumn{2}{|c|}{Tables \ref{CL results accident 1B}, \ref{CL results accident 2A} and \ref{CL results accident 2B}}& \multicolumn{2}{|c|}{Bootstrap}\\
  & True OLL$^\ddagger$ & RBNS  &Error$^\ddagger$& Mean &   Est.Err. \\
\hline\hline
Cumulative payments, Listing \ref{CLCode2} &19,735	&19,076	&-658& 19,000 & 942\\		
Cumulative payments and claim status, Listing \ref{CLCode3} &19,735&20,080&346&19,998&955\\
Cumulative payments and all covariates, formula \eqref{linear regression number 2} &19,735	&20,108	&374&20,020&963\\	
  \hline
\end{tabular}
\end{center}}
\caption{Bootstrap results (aggregated all claims of all accident years) of the linear regression models using different sets of covariates according to Tables \ref{CL results accident 1B}, \ref{CL results accident 2A} and \ref{CL results accident 2B}.}
\label{Results accident 3A}
\end{table}

Completely analogously to Section \ref{Lab: chain-ladder reserving and individual bootstrap}, we perform an individual claims history bootstrap analysis, resampling the upper individual claims triangle by drawing with replacement. The selected individual claims are used to compute the bootstrap estimates $\widehat{\mu}^\ast_{j-1}$ of the
three regression functions given in \eqref{weighted square loss 2} (only individual cumulative payments; Table \ref{CL results accident 1B}), 
\eqref{linear regression claim status} (individual cumulative payments and claim status; Table \ref{CL results accident 2A}) and
\eqref{linear regression number 2} (all covariates; Table \ref{CL results accident 2B}). These bootstrapped regression functions are then used to complete the lower triangle on the originally observed claims, i.e., similar to \eqref{bootstrap extrapolation} we extrapolate the real observed upper triangle with the bootstrapped PtU factors. We perform this over 1,000 bootstrap samples (each having the same sample size as the original upper individual claims triangle). The aggregated results over all claims are presented in Table \ref{Results accident 3A}, and they are compared to the non-bootstrapped results of Table \ref{CL results accident 1B}, \ref{CL results accident 2A} and \ref{CL results accident 2B}.
We give the following remarks on Table \ref{Results accident 3A}:
\begin{itemize}
\item The original RBNS reserves and the bootstrap means are very close (in all three cases the difference is roughly 80). This indicates consistency in the sense that the bootstrap does not collect a major bias.
\item The bootstrap 'Est.Err.' corresponds to the standard deviation in the bootstrapped ultimate claim predictions (aggregated over all claims). This can be interpreted as the average model estimation error, similar to the estimation error in Mack's \cite{Mack} RMSEP formula, see Section \ref{Lab: chain-ladder reserving and individual bootstrap}. This model uncertainty estimate has a similar magnitude as in Table \ref{CL results}, indicating that the impact of IBNR claims is negligible in this example on model estimation error (this will be different in the example in Section \ref{The role of claims incurred}, below). Second, we observe from Table \ref{Results accident 3A} that this error is slightly increasing with model complexity. Thus, more model complexity seems to increase estimation uncertainty in this example.
\item The 'Est.Err.' only accounts for model error and not for process variance (irreducible risk). Nevertheless, the numbers of 'Est.Err.' in Table \ref{Results accident 3A} dominate the observed forecast errors `Error$^\ddagger$', which implies that these linear regression models cannot be rejected for individual claims RBNS forecasting.
\end{itemize}

\begin{table}[h]
\centering
{\footnotesize
\begin{center}
\begin{tabular}{|l|rrrr||rrrr|}
\hline
& \multicolumn{4}{|c||}{Linear regression model \eqref{linear regression claim status}; Table \ref{CL results accident 2A}}
& \multicolumn{4}{|c|}{Linear regression model \eqref{linear regression number 2}; Table \ref{CL results accident 2B}}\\
&	Ind.RMSE$^\ddagger$ & Ind.RMSE$^\ddagger$	& & 
Bootstrap&	Ind.RMSE$^\ddagger$ & Ind.RMSE$^\ddagger$	& & 
Bootstrap \\
$i$& Listing \ref{CLCode} & Listing \ref{CLCode3} &Difference& Est.Err.
& Listing \ref{CLCode} & All covariates &Difference& Est.Err.
\\\hline\hline
2&	1.499&	1.455&	-0.044&	0.031&	1.499&	1.455&	-0.044&	0.048\\
3&	2.956&	3.012&	0.056&	0.049&	2.956&	3.013&	0.057&	0.086\\
4&	4.263&	4.221&	-0.042&	0.058&	4.263&	4.221&	-0.042&	0.110\\
5&	8.240&	8.135&	-0.105&	0.077&	8.240&	8.121&	-0.119&	0.154\\
 \hline
\end{tabular}
\end{center}}
\caption{Bootstrap results on individual claims in the two linear regression models
\eqref{linear regression claim status} (payments and claim status) and \eqref{linear regression number 2} (all covariates), see also 
 Tables \ref{CL results accident 2A} and \ref{CL results accident 2B}.}
\label{Results accident 3B}
\end{table}

We can also analyze the bootstrap results on an individual claims level.
Table \ref{Results accident 3B} presents the results per accident year.
The individual RMSE `Ind.RMSE$^\ddagger$' decreases in accident year $i=5$ from the model of Listing \ref{CLCode} of 8.240 to 8.135 for the model of Listing \ref{CLCode3}. This is a decrease of -0.105, the bootstrap standard deviation for this quantity accounting for model uncertainty is 0.077. Thus, in this example the decrease exceeds the size of the model uncertainty. On the other hand, we observe that the value of 0.077 is more than 100 times smaller than the individual RMSE of 
8.135. Not surprisingly, this shows that the driver of individual claims uncertainty is irreducible risk, i.e., we are in a typical situation of a low signal-to-noise ratio, and we do not expect very accurate reserves on individual claims, but only on aggregated claims, e.g., aggregated within the accident periods, see Figure \ref{figure reserves open claims 2}. This low signal-to-noise situation can only be improved by better covariates, e.g., claims incurred or medical reports may be useful to provide more accurate forecasts on individual claims, for instance, making a statement about the expected recovery time after an accident.

\subsection{Lab: Accident insurance example -- feed-forward neural network}
\begin{tcolorbox}[title=]
This section replaces the linear regression model of Section \ref{Lab: Accident insurance example -- linear regression}  by a feed-forward neural network, still making the Markov assumption on the input covariates.
\end{tcolorbox}

\label{Lab: Accident insurance example -- feed-forward neural network}
The results of Table \ref{CL results accident 2B} are based on linear regressions \eqref{linear regression number 2}. We replace these linear regressions by a feed-forward neural network (FNN) architectures $\mu^{\rm FNN}_{j-1}$, for $1\le j\le J$, allowing for more modeling flexibility, capturing non-linear terms and allowing for more complex interactions between the covariate components. The specific selected FNN architecture is documented in Table \ref{architecture table} and the full code is given in Listing \ref{FNNCode} in the appendix.

\begin{table}[h]
\centering
{\footnotesize
\begin{center}
\begin{tabular}{|l||c|c|c|}
\hline
Module & Dimension & $\#$ Weights & Activation
\\\hline\hline
Input layer & 6 & --& -- \\
1st hidden layer & 20 & 140 & GELU \\
2nd hidden layer & 15 & 315 & GELU \\
Output layer & 1 & 16 & identity \\
\hline
\end{tabular}
\end{center}}
\caption{Selected FNN architectures $\mu^{\rm FNN}_{j-1}$, for $1\le j\le J$, in the accident insurance example.}
\label{architecture table}
\end{table}

The remaining modeling parts are very similar to Section \ref{Lab: Accident insurance example -- linear regression}, only the linear regression part
\eqref{linear regression number 2} is replaced by the FNNs $\mu^{\rm FNN}_{j-1}$, for $1\le j\le J$. We use the same covariates, but the accident month enters as a continuous variable and we manually add the interaction term between the cumulative payments and the claim status. The specifications of the stochastic gradient descent fitting procedure are provided in Table \ref{hyperparameters}, see also Listing \ref{FNNCode} in the appendix.

\begin{table}[h]
\footnotesize
\centering
\begin{tabular}{ll}
\toprule
\textbf{Component} & \textbf{Setting} \\
\midrule
Loss function & mean squared error (MSE)\\
Optimizer & Adam with learning rate $10^{-3}$ \\
Batch size and epochs & 8,192 and 500\\
Learning-validation split & $9:1$\\
Early stopping & reduce learning rate on plateau, factor 0.9, patience 5\\
Ensembling & 10 network fits with different seeds \\
\bottomrule
\end{tabular}
\caption{Key implementation and hyper-parameters for FNN fitting.}
\label{hyperparameters}
\end{table}

\medskip

There is one key feature that is worth mentioning, we refer to line {\tt 58} of Listing \ref{FNNCode}. Namely, the linear regression model using the square loss function provides an estimated solution $\widehat{\mu}_{j-1}$ that satisfies the balance property, i.e., for the MLE estimated linear regression we have
\begin{equation}\label{balance property}
\sum_{i=1}^{I-j}
\sum_{\nu:\, T_{i|\nu} \le j-1}
\widehat{\mu}_{j-1}\left((C_{i,l|\nu}, \bX_{i,l|\nu})_{l=0}^{j-1}\right)
=
\sum_{i=1}^{I-j}
\sum_{\nu:\, T_{i|\nu} \le j-1}
\widehat{C}_{i,J|\nu}.
\end{equation}
This is a consequence of working with the canonical link under the square loss function (in an exponential dispersion family (EDF) setting), and it says that the average estimated model is equal to the average response \eqref{balance property}; see also Lindholm--W\"uthrich \cite[Proposition 2.6]{LindholmW}. This is an in-sample unbiasedness property, and it implies that there are no (obvious) biases that can propagate through the recursive structure (of course, assumed we have stationarity along the accident period axis). Unfortunately, stochastic gradient descent (SGD) fitted models fail to satisfy this balance property. Therefore, we need to enforce it by a post calibration step
\begin{equation}\label{post calibration}
\widehat{\mu}^{\rm FNN}_{j-1}(\cdot) \quad
\longleftarrow \quad
\frac{\sum_{i=1}^{I-j}
\sum_{\nu:\, T_{i|\nu} \le j-1}
\widehat{C}_{i,J|\nu}}{\sum_{i=1}^{I-j}
\sum_{\nu:\, T_{i|\nu} \le j-1}
\widehat{\mu}^{\rm FNN}_{j-1}\left((C_{i,l|\nu}, \bX_{i,l|\nu})_{l=0}^{j-1}\right)}~\widehat{\mu}^{\rm FNN}_{j-1}(\cdot).
\end{equation}
That is, we apply a multiplicative scaling step to enforce the balance property
\eqref{balance property} by the new (scaled) regression function (one could also shift the intercept correspondingly).
This post calibration step \eqref{post calibration} helps to control a potential bias, as we now have an in-sample unbiased model, for the code see line {\tt 58} of Listing \ref{FNNCode}.

\begin{table}[h]
\centering
{\footnotesize
\begin{center}
\begin{tabular}{|c|r|rr|rr|rr|}
\hline
 &  & RBNS  & RBNS   &  Error$^\ddagger$ &Error$^\ddagger$ & Ind.RMSE$^\ddagger$& Ind.RMSE$^\ddagger$  \\
$i$ & True OLL$^\ddagger$ & Listing \ref{CLCode} & FNN all cov. &  Listing \ref{CLCode} & FNN all cov.&  Listing \ref{CLCode} & FNN all cov.  \\
\hline\hline
1 &0	&0	&0	&0	&0&0&0	\\
2&353&	339&	512&	-14	&	159	&	1.499&	\bl{1.491}\\
3 &1,017&	1,305&	1,500&	288&484&2.956&\bl{3.017}\\
4 &3,102	&3,099	&3,399&-2&297&	4.263&	\bl{4.218}\\
5 &15,263&	14,216	&15,395	&-1,046&132&	8.240&\bl{8.114}\\\hline
Total &19,735	&18,959	&20,806	&-774&1,072&&\\		
 \hline
\end{tabular}
\end{center}}
\caption{Accident insurance example following up Table \ref{CL results accident 2B}: RBNS results of individual claims prediction using  Listing \ref{CLCode} and a FNN architecture on all available covariates of settlement period $j-1$;
 the earmarked columns$^\ddagger$ use the ground truth
in the lower triangle.}
\label{CL results accident 2C}
\end{table}

Table \ref{CL results accident 2C} presents the results of the FNN architectures and they should be compared to the linear regression results of Table 
\ref{CL results accident 2B}. The conclusion is simple, the older accident years $i=2,3$ do not benefit from the additional modeling flexibility, mainly because SGD fitting is not as efficient as Fisher's scoring method to fit a linear regression/GLM. In fact, the out-of-sample validation control triggering early stopping lets the older accident years be worse than in the simple linear regression model. The more recent accident years $i=4,5$ marginally improve compared to the linear regression model, see Tables \ref{CL results accident 2B} and \ref{CL results accident 2C} (\bl{blue colors}). Here, we benefit from more flexible functional forms compared to the linear regression. However, the improvement is comparably minor, and in view of the computational efficiency (and the explainability) of the linear regression, we give preference to the linear regression model in this accident insurance example.

Naturally, at this stage we could also exploit other ML methods such as gradient boosting machines (GBMs). For the moment, we refrain from doing so.

\subsection{Transformer architecture}
\label{Transformer architecture}
\begin{tcolorbox}[title=]
In the last step of the accident insurance example, we lift the regression model to a transformer architecture being able to process the entire past claims history, i.e., we drop the Markov assumption on the input used in the previous section.
\end{tcolorbox}

The natural next step is to replace the FNN architecture (used in the previous section) by a transformer architecture that allows one to use the entire past claim history
\begin{equation*}
(C_{i,l|\nu}, \bX_{i,l|\nu})_{l=0}^{j-1}
  ~\mapsto ~
  \mu^{\rm transf}_{j-1}\left((C_{i,l|\nu}, \bX_{i,l|\nu})_{l=0}^{j-1}\right).
\end{equation*}
Listing \ref{TransformerCode} in the appendix gives the code that we have used to compute the next example (the listing focuses on the differences to Listing \ref{FNNCode}); we mention that in this transformer architecture we only select ``simple'' linear embeddings, but this approach could easily accommodate more complex functional forms.
The transformer architecture can be applied to all periods $j-1=1,\ldots, J-1$. For $j-1=0$, we have only one observed past period, and we therefore use the FNN architecture of the previous section. Since there are only the two stochastic dynamic covariates of cumulative payments and claim status, we restrict our next example to these two stochastic processes
\begin{equation*}
(C_{i,l|\nu}, O_{i,l|\nu})_{l=0}^{j-1}
  ~\mapsto ~
  \mu^{\rm transf}_{j-1}\left((C_{i,l|\nu}, O_{i,l|\nu})_{l=0}^{j-1}\right),
\end{equation*}
and the results of Table \ref{CL results accident transformer} should be compared to Table \ref{CL results accident 2A}.

\begin{table}[h]
\centering
{\footnotesize
\begin{center}
\begin{tabular}{|c|r|rr|rr|rr|}
\hline
 &  & RBNS  & RBNS   &  Error$^\ddagger$ &Error$^\ddagger$ & Ind.RMSE$^\ddagger$& Ind.RMSE$^\ddagger$  \\
$i$ & True OLL$^\ddagger$ & Listing \ref{CLCode} & Transformer &  Listing \ref{CLCode} & Transformer&  Listing \ref{CLCode} & Transformer  \\
\hline\hline
1 &0	&0	&0	&0	&0&0&0	\\
2&353&	339&	296&	-14	&	-57	&	1.499&	\bl{1.533}\\
3 &1,017&	1,305&	1,338&	288&322&2.956&\bl{3.004}\\
4 &3,102	&3,099	&3,260&-2&158&	4.263&	\bl{4.241}\\
5 &15,263&	14,216	&14,961	&-1,046&-301&	8.240&\bl{8.129}\\\hline
Total &19,735	&18,959	&19,855	&-774&122&&\\		
 \hline
\end{tabular}
\end{center}}
\caption{Accident insurance: RBNS results of individual claims prediction using  Listings \ref{CLCode} and a transformer architecture considering the cumulative payments and claim status history
$(C_{i,l|\nu}, O_{i,l|\nu})_{l=0}^{j-1}$; the earmarked columns$^\ddagger$ use the ground truth
in the lower triangle.}
\label{CL results accident transformer}
\end{table}

Comparing the results of Table \ref{CL results accident 2A} and 
Table \ref{CL results accident transformer}, we conclude that the additional model complexity is not fully justified in our forecast problem. This is likely because we have a rather small dataset ($5\times 5$ triangle) on a comparable coarse time grid. For instance, for accident year $i=2$, this implies that the input time-series has a total length of 4, i.e., this is not a typically length a transformer architecture brings major benefits. Thus, we have technically verified that this set-up can be implemented and computed, the proof whether it is beneficial to increase predictive performance on bigger datasets still needs to be done.

\section{The role of claims incurred}
\label{The role of claims incurred}
\begin{tcolorbox}[title=] 
This section presents our second example where in addition to individual cumulative payments and the claim status process also individual claims incurred information is available. We study different models to evaluate the explanatory power of these different inputs.
\end{tcolorbox}

The results in Section \ref{Lab: Accident insurance example -- linear regression} have highlighted the importance of the claim status process $O_{i, 0:J|\nu}$ for forecasting ultimate claims. This section analyzes the role of claims incurred $I_{i, 0:J|\nu}$ which are individual case estimates set by claims adjusters. For this we consider our second example introduced in Table \ref{tab:liablitydata}. 
We build a linear regression model including individual cumulative payments $C_{i,j-1|\nu}$, claims incurred $I_{i,j-1|\nu}$ and the claim status $O_{i,j-1|\nu}$ of the latest period $j-1$
\begin{eqnarray}\nonumber
\mu_{j-1}\left((C_{i,l|\nu}, \bX_{i,l|\nu})_{l=0}^{j-1}\right)
&=& \vartheta_0 + \vartheta_1 \,C_{i,j-1|\nu}+ 
\vartheta_2 \,I_{i,j-1|\nu}+ \vartheta_3 \,O_{i,j-1|\nu}+ 
\\&&+ ~\label{linear regression claim status and claims incurred}
\vartheta_4\, C_{i,j-1|\nu}\,O_{i,j-1|\nu}
+ \vartheta_5\, I_{i,j-1|\nu}\,O_{i,j-1|\nu}.
\end{eqnarray}
This model considers linear terms in individual cumulative payments, claims incurred and claim status, and we also let the claim status interact with the other two inputs.

\begin{table}[h]
\centering
{\footnotesize
\begin{center}
\begin{tabular}{|c||c|c|c|c|c|}
\hline
Model & $C_{i,j-1|\nu}$ &$I_{i,j-1|\nu}$ &$O_{i,j-1|\nu}$ & $C_{i,j-1|\nu}\,O_{i,j-1|\nu}$ &$ I_{i,j-1|\nu}\,O_{i,j-1|\nu}$\\ 
 \hline\hline
 Model C & x &&&&\\\hline
 Model I &  &x&&&\\\hline
 Model CO & x &&x&x&\\\hline
 Model IO &  &x&x&&x\\\hline
 Model CIO & x &x&x&x&x\\
 \hline
\end{tabular}
\end{center}}
\caption{Liability insurance: RBNS models considering different versions of \eqref{linear regression claim status and claims incurred}.}
\label{incurred}
\end{table}

We consider five different versions of the linear regression function
\eqref{linear regression claim status and claims incurred} by excluding selected terms. The five considered variants are illustrated in Table \ref{incurred}, the final Model CIO includes all the terms. Each of these models is fitted and we compute the resulting
individual claims RMSEs 'Ind.RMSE$^\ddagger$' measuring the individual claim forecast against the ground truth individual OLL, see \eqref{individual RMSE}.

\begin{table}[h]
\centering
{\footnotesize
\begin{center}
\begin{tabular}{|c||r|r ||r|r||r||r|}
  \hline
  & \multicolumn{5}{|c||}{Linear regression \eqref{linear regression claim status and claims incurred}}&
  \multicolumn{1}{|c|}{FNN}\\
 $i$& Model C & Model I & Model CO & Model IO & Model CIO & Model CIO  \\
\hline\hline
1&	0&	0&	0&	0&	0 & 0 \\
2&	2.628&	10.066&	2.612&	4.781&	2.571& 2.633\\
3&	19.964&	16.748&	19.794&	16.481&	17.749&17.076\\
4&	12.489&	9.791&	12.339&	8.559	&8.794&8.510\\
5&	14.290&	14.402&	14.268&	14.138&	13.872& 13.786\\
 \hline
\end{tabular}
\end{center}}
\caption{Liability insurance: Individual claims RMSEs `Ind.RMSE$^\ddagger$', see \eqref{individual RMSE}.}
\label{incurred 1}
\end{table}

The results are presented in Table \ref{incurred 1} and we give the following remarks.
\begin{itemize}
\item Model C and Model I: We observe that the claims incurred $I_{i,j-1|\nu}$ seems to have superior predictive power compared to individual cumulative payments $C_{i,j-1|\nu}$, except in accident year $i=2$. A reason for the different behavior in this old accident year may be that the claims incurred estimates have not been continuously updated by the claims adjusters for claims close to settlement. In that case, the payments made give a more accurate forecast.
\item Model CO and Model IO: In combination with the claim status information $O_{i,j-1|\nu}$, we give preference to the claims incurred information giving more accurate forecasts than the individual cumulative claim version. Again only for the accident year $i=2$, Model IO does not outperform Model CO, however, the gap has decreased.
  \item Model CIO: If we combine the two models to Model CIO, we receive a generally strong model, though not in all accident years the best one on individual claims. This indicates that we should include all information, but it also seems that the linear regression structure can be improved. This is verified by the last column where we replace the linear regression models \eqref{linear regression claim status and claims incurred} by FNNs $\mu_{j-1}^{\rm FNN}$ on the identical input information; for the FNN architecture see also Listing \ref{FNNCode} in the appendix.
\end{itemize}

\begin{table}[h]
\centering
{\footnotesize
\begin{center}
\begin{tabular}{|c|r|rr|rr|rrr|}
\hline
 &  & RBNS  & RBNS   &  Error$^\ddagger$ &Error$^\ddagger$ & Ind.RMSE$^\ddagger$& Ind.RMSE$^\ddagger$ & Bootstrap \\
$i$ & True OLL$^\ddagger$ & Listing \ref{CLCode} & Model CIO &  Listing \ref{CLCode} & Model CIO&  Listing \ref{CLCode} & Model CIO & Est.Err. \\
\hline\hline
1 &0	&0	&0	&0	&0&0&0	& \\
2&361&	635&	442&	274	&	81	&	2.717&	\bl{2.571}&0.061\\
3 &3,233&1,497&	1,398&	-1,736&-1,835&19.988&\bl{17.749}&0.119\\
4 &3,287&2,488	&2,938&-799&-349&	12.400&	\bl{8.794}&0.156\\
5 &4,613&3,982	&4,172	&-631&-440&	14.901&\bl{13.872}& 0.216\\\hline
Total &11,494	&8,601	&8,950	&-2,893&-2,543&&&\\		
  \hline
  \multicolumn{5}{|l|}{Bootstrap Est.Err. (aggregated claim)}&886&&&\\
  \hline
\end{tabular}
\end{center}}
\caption{Liability insurance: RBNS results of individual claims prediction using  Listings \ref{CLCode} and
the linear regression Model CIO considering 
$(C_{i,j-1|\nu}, I_{i,j-1|\nu}, O_{i,j-1|\nu})$; the earmarked columns$^\ddagger$ use the ground truth
in the lower triangle.}
\label{incurred 2}
\end{table}

We come back to the RBNS CL predictions given in Table \ref{CL results RBNS and IBNR} for the liability insurance data set, and we complement these results with the individual claims reserving results of Table \ref{incurred 1} -- we select the linear regression model \eqref{linear regression claim status and claims incurred} called Model CIO. Moreover, we perform an individual claims history bootstrap analysis as described in Section \ref{Lab: Linear regression bootstrap results} -- this can be done because linear regression fitting is very fast. We interpret the results of Table \ref{incurred 2}:
\begin{itemize}
\item The linear regression Model CIO generally improves the results compared to the RBNS CL of Table \ref{CL results RBNS and IBNR}, saying that the combination of individual cumulative payments, claims incurred and claim status is beneficial to improve forecast accuracy. This is also verified by the individual claims RMSEs in columns 'Ind.RMSE$^\ddagger$'.
\item There is a severe under-estimation in accident year $i=3$. This under-estimation can be traced back to two individual claims that became very large in development periods $j=3,4$, we have already documented this in \cite{PtU}. These two 'outliers' also explain the large value in the individual claims RMSE 'Ind.RMSE$^\ddagger$'. Thus, the model could not capture these two strongly increasing claims (amounting to payments of 1,874), but apart from that the forecasts look very good. PS: These two large claims should not be called `outliers' because they are not data error, but real claims that need to be paid by the insurer.
\item The bootstrap analysis provides an overall model estimation uncertainty of 886, which looks reasonable and adding the irreducible risk (not explicitly assessed here) explains the forecast error 'Error$^\ddagger$' of -2,543 (this includes the two large claims of accident year $i=3$).
\item The last column of Table \ref{incurred 2} gives the bootstrap estimation uncertainty on individual claims. Similar to Table \ref{Results accident 3B}, we conclude that the by far most dominant term is irreducible risk (low signal-to-noise ratio), but the gap is a bit smaller in this liability insurance example compared to the accident insurance example in Table \ref{Results accident 3B}, which may be explained by the additional claims incurred information.
  \end{itemize}

\section{IBNR reserving}
\label{sec: IBNR reserving}

\begin{tcolorbox}[title=] 
The last missing piece is to compute the IBNR reserves for the claims not reported yet at the evaluation date $I$.
\end{tcolorbox}

The last part of the reserving exercise is to predict the IBNR claims. These are not included in the previously computed RBNS reserves.
There are many different ways to do so, and often a frequency-severity model is proposed, see, e.g., Parodi \cite{Parodi}. The first modeling part of the frequency-severity model predicts the number of IBNR claims, which can be seen as a reporting delay censoring problem. Popular methods for predicting these counts either use aggregate CL type methods or they use methods from survival analysis. This will result in a reporting pattern of the total number of claims $N_i$ occurred in period $i$, which allows one to predict the number of late reportings. This analysis can also involve an exposure measure, such as premium earned, and additional risk factor information. For the severities, one then studies a cross-classified model having the accident date on one axis and the reporting delay on the other axis. Using the RBNS predictions $\widehat{C}^{\rm RBNS}_{i,J|\nu}$ together with their claim's reporting delays $T_{i|\nu}$ allows one to predict the sizes of the late reported claims in such a cross-classified model. One can further refine this by contract and claim feature information which results in a proposal similar to the one in the addendum to Semenovich \cite{Semenovich}.

We take a simpler approach which is still very accurate for our data. We directly estimate the IBNR amounts with a cross-classified CL model without going through the frequency-severity split. Obviously, this uses less granular data. Consider all RBNS claim predictions $\widehat{C}^{\rm RBNS}_{i,J|\nu}$ of the claims $\nu$ being reported by the evaluation date $I$, i.e., with $i+T_{i|\nu} \le I$. This concerns the claims reported in the upper triangle with ultimate claim forecasts obtained, e.g., by Algorithm \ref{algorithm: PtU one-shot}.
Based on this, we build a new upper triangle given by defining the entries
\begin{equation}\label{Upper RBNS triangle}
  S_{i,j} 
 = \sum_{\nu=1}^{N_i}\widehat{C}^{\rm RBNS}_{i,J|\nu} \,\mathds{1}_{\{T_{i|\nu} =j \}}\,
  = \sum_{\nu:\, T_{i|\nu} =j} \widehat{C}^{\rm RBNS}_{i,J|\nu}
  \qquad \text{ for $i+j \le I$.}
\end{equation}
This is the total predicted claim amount of accident period $i$ that has been reported with a reporting lag of $j$. If we wanted to build a frequency-severity model, we would divide this by the observed number of reported claims with that reporting lag, i.e., 
\begin{equation*}
N_{i,j}=  \sum_{\nu=1}^{N_i}\mathds{1}_{\{T_{i|\nu} =j \}}\,
=\sum_{\nu:\, T_{i|\nu} =j} 1.
\end{equation*}
 However, for the results below we directly use the data (upper triangle)
\begin{equation*}
  {\cal S}_I = \left\{ S_{i,j};~   i+j \le I,\, 1\le i \le I,\, 0\le j \le J \right\},
\end{equation*}
and the lower (IBNR) triangle at time $I$ is forecasted with a simple CL prediction.

\begin{table}[h]
\centering
{\footnotesize
\begin{center}
\begin{tabular}{|l|r|rr|rr|}
\hline
 & True OLL$^\ddagger$ & Reserves& CL RMSEP & Error$^\ddagger$ &  \% CL RMSEP$^\ddagger$\\
\hline\hline
\underline{Accident dataset} &&&&&\\
Mack's CL model \cite{Mack} & 24,212&	23,064&1,663&-1,148 &69\%\\
RBNS CL prediction of Table \ref{CL results RBNS and IBNR} & 19,735&	18,959&--&-774 &--\\
  IBNR CL prediction of Table \ref{CL results RBNS and IBNR} & 4,478&	4,105&--&-374 &--\\
\hline
Total (of next two lines)  & 24,212&	\bl{24,430}&--&217 &13\%\\
Individual RBNS of Table \ref{CL results accident 2B} & 19,735&	\bl{20,108}&--&374 &--\\
IBNR reserving using \eqref{Upper RBNS triangle}& 4,478&	\bl{4,322}&--&-156 &--\\  
 \hline\hline
\underline{Liability dataset} &&&&&\\
Mack's CL model \cite{Mack} & 15,730&	11,526&	1,977&	-4,204&	213\%\\
RBNS CL prediction of Table \ref{CL results RBNS and IBNR} & 11,494&	8,601&--&-2,893 &--\\
  IBNR CL prediction of Table \ref{CL results RBNS and IBNR} & 4,236&	2,925&--&-1,311 &--\\
\hline
Total (of next two lines)  & 15,730&	\bl{12,486}&--&-3,244 &164\%\\
Individual RBNS of Table \ref{incurred 2} & 11,494&	\bl{8,950}&--&-2,543 &--\\
IBNR reserving using \eqref{Upper RBNS triangle}& 4,236&	\bl{3,536}&--&-700 &--\\    
 \hline
\end{tabular}
\end{center}}
\caption{Mack's CL results on cumulative payments split to RBNS and IBNR reserves; the earmarked columns$^\ddagger$ can only be computed because we know the lower triangle in our examples.}
\label{CL results RBNS and IBNR Individidual}
\end{table}

\medskip

The results are presented in Table \ref{CL results RBNS and IBNR Individidual}, and they are compared to the CL analysis of Table \ref{CL results RBNS and IBNR}. From Table \ref{CL results RBNS and IBNR Individidual} we observe that in our examples we get IBNR reserves that are very accurate, i.e., more accurate than the ones of Table 
\ref{CL results RBNS and IBNR}. In the accident insurance example we lower the prediction error from -374 to -156, and in the liability insurance example from -1,311 to -700. Thus, this simple method performs very well on these two (small-scale) datasets. This also impacts the total RBNS + IBNR reserves, being more accurate than in Mack's CL model. This completes our numerical examples.


\section{Summary}
\label{Conclusions and Outlook}


Building on our previous paper \cite{PtU}, we introduced several refinements to the one-shot estimation and prediction procedure based on individual claims histories. An exciting observation in our examples was that linear regressions perform quite well in this one-shot forecasting problem. Since linear regressions can be fitted very fast, this moreover allows one to perform an individual claims history bootstrap to assess model estimation uncertainty.

Our examples are small-scale examples in the sense that they use $5\times 5$ years observations, and it remains an open question to verify that our proposal also works on bigger data. Another open point is to take care of non-stationarity, e.g., caused by inflation. In our examples, a simple balance property step was sufficient. However, in other situations manual interventions may be necessary to cope with non-stationarity.

\begin{itemize}
\item In a next step, bigger data should be studied, and also the impact of longer time-series inputs for forecasting ultimate claims needs to be understood, e.g., using transformer architectures.
\item In our examples, an additive regression structure seems to be better than a multiplicative one. The deeper reason for this preference is not entirely clear. Also the role of the claims that are precisely zero needs to be explored, because neither in the additive nor in the multiplicative setting, these can easily be modeled/fitted.
\item The one-shot ultimate claim prediction can be complemented by a cash flow pattern for the RBNS reserves, e.g., using a transformer decoder architecture.
\item We used one of the most simple approaches to predict IBNR claims. Certainly there are many different ways to enhance this procedure and estimate.
\end{itemize}

\bigskip

{\small 
\renewcommand{\baselinestretch}{.51}
}

\newpage

\appendix


\begin{lstlisting}[float=h,frame=tb,caption={Recursive one-shot PtU RBNS algorithm: FNN regression.}, label=FNNCode]
### pre-processing covariates
m1 <- mean(as.matrix(triCC[,paste0("X", 0:J0)]), na.rm=TRUE)
s1 <- sd(as.matrix(triCC[,paste0("X", 0:J0)]), na.rm=TRUE)
triCC[,paste0("X", 0:J0)] <- (triCC[,paste0("X", 0:J0)]-m1)/s1
claims$AccMonth           <- (claims$AccMonth-1)/11
claims$WorkLeisure        <- as.integer(claims$WorkLeisure)-1
claims$RepDelay           <- pmin(365, claims$RepDelayDays)/365

### network architecture
FNN <- function(seed, q0){
    tf$keras$backend$clear_session(); set.seed(seed); set_random_seed(seed)
    Design   <- layer_input(shape=c(q0[1]), dtype='float32')
    Network = Design %>% layer_dense(units=q0[2], activation='gelu') %>%
                         layer_dense(units=q0[3], activation='gelu') %>%
                         layer_dense(units=1, activation='linear')
    keras_model(inputs=c(Design), outputs=c(Network))
    }

### initialize ultimate claims with observed ones
claims$YY <- NA
claims[which(claims$AccDate<=I0-J0),]$YY <- claims[which(claims$AccDate<=I0-J0),]$Ultimate

### recursive network fitting and PtU forecasting
for (j in rev(0:(J0-1))){
    i <- I0-j-1
    ### prepare learning and forecast data for FNN
    select  <- which((claims$AccDate<=i)&(claims$RepDelayYY<=j))
    Ylearn0 <- as.matrix(claims[select,]$YY)
    mu.hom  <- mean(Ylearn0)
    mu.sd   <- sd(Ylearn0)
    Ylearn  <- (Ylearn0-mu.hom)/mu.sd
    Xlearn  <- as.matrix(cbind(triOO[select,paste0("X",j)],
                               triCC[select,paste0("X",j)]*triOO[select,paste0("X",j)],
                               claims[select,"WorkLeisure"],
                               claims[select,"RepDelay"],
                               claims[select,"AccMonth"],
                               triCC[select,paste0("X",j)]))
    select  <- which(claims$AccDate==(i+1))
    Xtest   <- as.matrix(cbind(triOO[select,paste0("X",j)],
                               triCC[select,paste0("X",j)]*triOO[select,paste0("X",j)],
                               claims[select,"WorkLeisure"],
                               claims[select,"RepDelay"],
                               claims[select,"AccMonthXX"],
                               triCC[select,paste0("X",j)]))
    ### network fitting
    model <- FNN(seed, c(ncol(Xlearn), c(20,15)))
    adam = optimizer_adam(learning_rate=0.001)
    model %>% compile(optimizer=adam, loss="mse")
    path1 <- paste0("./Networks/FNN_",seed,"_j",j,".weights.h5")
    model_write = callback_model_checkpoint(path1, save_best_only=T, save_weights_only=T)
    learn_rate = callback_reduce_lr_on_plateau(factor=0.9, patience=5, cooldown=0)
    fit <- model %>% fit(list(Xlearn), Ylearn, validation_split=0.1, batch_size=8192, 
                         epochs=500, callbacks=list(model_write, learn_rate), shuffle=TRUE)
    ### results
    model$load_weights(path1)
    learn.NN <- mu.hom + mu.sd * model %>% predict(list(Xlearn), batch_size=10^6, verbose=0)
    test.NN <-  mu.hom + mu.sd * model %>% predict(list(Xtest), batch_size=10^6, verbose=0)
    claims[which(claims$AccDate==(i+1)),]$YY <- sum(Ylearn0)/sum(learn.NN) * test.NN
 }
\end{lstlisting}

\begin{lstlisting}[float=h,frame=tb,caption={Recursive one-shot PtU RBNS algorithm: Transformer regression.}, label=TransformerCode]
### network architecture
Transformer <- function(seed, input_size, units0){
  tf$keras$backend$clear_session(); set.seed(seed); set_random_seed(seed)
  Design <- layer_input(shape=c(input_size[1], input_size[2]), dtype='float32')
  #
  Repr   <- Design %>% time_distributed(layer_dense(units=units0[1]))
  query  <- Repr %>% time_distributed(layer_dense(units=units0[1]))
  key    <- Repr %>% time_distributed(layer_dense(units=units0[1]))
  value  <- Repr %>% time_distributed(layer_dense(units=units0[1]))
  #
  attention_output <- list(query, value, key) %>% layer_attention(use_scale=TRUE, trainabl=TRUE)
  #
  skip_1 <- layer_add(list(attention_output, Repr))
  skip_2 <- skip_1 %>% layer_layer_normalization() %>% 
                       time_distributed(layer_dense(units=units0[1]))
  #
  Features <- layer_add(list(skip_2, skip_1)) %>% layer_flatten()
  #
  Network = Features %>% layer_dense(units=units0[2], activation='gelu') %>%
                         layer_dense(units=units0[3], activation='gelu') %>%
                         layer_dense(units=1, activation='linear')
  #
  keras_model(inputs = c(Design), outputs = c(Network))
}

### recursive network fitting and PtU forecasting
.
.
.
    ### learning data
    select  <- which((claims$AccDate<=i)&(claims$RepDelayYY<=j))
    Xlearn0 <- array(NA, dim=c(length(select), j+1, 3))
    Xlearn0[,,1] <- as.matrix(triOO.XX[select,paste0("X",0:j)])
    Xlearn0[,,2] <- as.matrix(triCC.XX[select,paste0("X",0:j)]*triOO.XX[select,paste0("X",0:j)])
    Xlearn0[,,3] <- as.matrix(triCC.XX[select,paste0("X",0:j)])
    ### forecast data
    select <- which(claims$AccDate==(i+1))
    Xtest <- array(NA, dim=c(length(select), j+1, 3))
    Xtest[,,1] <- as.matrix(triOO.XX[select,paste0("X",0:j)])
    Xtest[,,2] <- as.matrix(triCC.XX[select,paste0("X",0:j)]*triOO.XX[select,paste0("X",0:j)])
    Xtest[,,3] <- as.matrix(triCC.XX[select,paste0("X",0:j)])
 .
 .
 .
    input_size <- dim(Xlearn0)[2:3]
    units0 <- c(10, 15, 10)
    model <- Transformer(seed, input_size, units0)
   
 
\end{lstlisting}

\end{document}